\documentclass{article} 
\usepackage{epsfig} 
\usepackage{cite} 
\usepackage{amssymb} 
\usepackage[intlimits,reqno]{amsmath} 
\def\eqref#1{Eq.(\ref{#1})}

\def\oo{\infty} 
\newcommand{\labbel}[1]{\label{eq:#1}} 
\newcommand{\dnk}[1]{ {\mathfrak D}^dk_{#1} } 
\def\e{\hbox{e}} 
\def\Li{\hbox{Li}} 
\def\Cl{\hbox{Cl}} 
\def\Clpt{\hbox{Cl}_2\left(\frac{\pi}{3}\right)} 
\def\Ls{\hbox{Ls}} 
\def\Lstpvt{\hbox{Ls}_3\left(\frac{2\pi}{3}\right)} 
\def\ie{{\it i.e.}\phantom{.}} 
\def\iep{i\epsilon} 
\def\ru{\sqrt{u}} 

\begin{document} 

\begin{titlepage} 
\vspace*{-1cm} 
\begin{flushright} 
       \begin{minipage}{5cm} CERN-PH-TH/2004-089 \\ 
       \end{minipage} 
\end{flushright} 
\vskip 3.5cm 
\renewcommand{\topfraction}{0.9} 
\renewcommand{\textfraction}{0.0} 

\begin{center} 
\boldmath 
{\Large\bf Analytic treatment of the two loop equal mass sunrise graph 
}\unboldmath 
\vskip 1.cm 
{\large S. Laporta,}$^{b,a,}$ 
\footnote{{\tt e-mail: Stefano.Laporta@bo.infn.it}} 
{\large E. Remiddi\,}$^{e,c,a,}$ 
\footnote{{\tt e-mail: Ettore.Remiddi@bo.infn.it}} 
\vskip .7cm 
\vskip .4cm 
{\it 
  $^a$ INFN, Sezione di Bologna, I-40126 Bologna, Italy \\ 
  $^b$ Dipartimento di Fisica, Universit\`{a} di Parma, I-43100 Parma, 
       Italy \\ 
  $^c$ Dipartimento di Fisica, Universit\`{a} di Bologna, I-40126 Bologna, 
       Italy \\ 
  $^e$ Theory Division, CERN, CH-1211 Geneva 23, Switzerland \\ 
} 
\end{center} 
\vskip 2.6cm 
\begin{abstract} 
The two loop equal mass sunrise graph is considered in the continuous 
$d$-dimensional regularisation for arbitrary values of the momentum 
transfer. After recalling the equivalence of the expansions at $d=2$ 
and $d=4$, the second order differential equation for the scalar 
Master Integral is expanded in $(d-2)$ and solved by the variation 
of the constants method of Euler up to first order in $(d-2)$ included. 
That requires the knowledge of the two independent solutions of the 
associated homogeneous equation, which are found to be related to the 
complete elliptic integrals of the first kind of suitable arguments. 
The behaviour and expansions of all the solutions at all the singular points 
of the equation are exhaustively discussed and written down explicitly. 
\vskip .7cm 
{\it Key words}: Feynman diagrams, Multi-loop calculations 

{\it PACS}: 11.15.Bt, 12.20.Ds 
\end{abstract} 
\vfill 
\end{titlepage} 

\newpage 

\renewcommand{\theequation}{\mbox{\arabic{section}.\arabic{equation}}} 

\section{ Introduction. } 
\label{sec:intro} 
\setcounter{equation}{0} 
The two-loop sunrise selfmass graph with equal masses 
is not yet known analytically, despite its apparent simplicity. 
In this paper we fill that gap and evaluate it, for 
arbitrary values of the momentum transfer, in terms of a suitable 
set of new functions, whose analytic properties (in particular behaviours 
and expansions at all their potentially singular points) are worked out 
explicitly in full details. 
\par 
We follow the differential equation approach, already introduced 
in~\cite{CCLR1} in the arbitrary mass case for studying particular 
values of the momentum transfer, within the usual regularization scheme 
in $d$-continuous dimensions. We are interested, as usual, in the 
$d \to 4$ limit; but it is known from~\cite{Tarasov} how to relate 
algebraically the coefficients of the expansions at $d=4$ and those 
of the expansion at $d=2$, and as the expansion at $d=2$ turns out to be 
simpler, we discuss in fact the $d \to 2$ limit. 
\par 
In the equal mass limit the two loop sunrise has two Master Integrals 
(MI's) only, which satisfy a system of two first order differential equations 
in the momentum transfer $u$ (we will use often also the Euclidean variable 
$z=-u$, which is positive when $u$ is spacelike), 
equivalent to a single second order equation for the simpler 
of the two MI's. The equation, which is exact in $d$, when expanded 
around $d=2$ in powers of $(d-2)$ gives rise to a set of chained 
inhomogeneous equations all having the same homogeneous part; we solve 
them recursively by the variation of the constants method of Euler, 
which gives formally the solutions of the inhomogeneous equations in terms 
of the solutions of the common homogeneous equation. 
Only when the homogeneous solutions are known 
Euler's formula does provide an effective analytic 
evaluation; the bulk of the paper consists, indeed, in working out those 
homogeneous solutions. 
\par 
The paper is organized as follow. In Section~\ref{sec:diffeq} the 
differential equations are written; in Section~\ref{sec:expansions} 
the connection between the expansions at $d=4$ and at $d=2$ is 
established; in Section~\ref{sec:d=2} the expansion of the second 
order differential equation at $d=2$ is worked out. 
Section~\ref{sec:Building} discusses how the solutions of the homogeneous 
equation can be obtained; Section~\ref{sec:singpoints} presents the 
expansions in $z$ of the homogeneous solutions at all the singular 
points of the differential equation; Section~\ref{sec:interpol} 
discusses another set of functions, represented in the form of 
one-dimensional definite integral over a suitable parameter, which 
solve the homogeneous equation between any two nearby singular points 
(referred to as ``interpolating solutions"). Section~\ref{sec:homosol} 
then finally gives the proper analytic continuation of the homogeneous 
solutions for any value of $z$, and Section~\ref{sec:transfids} discusses 
the properties of the homogeneous equation and of its solutions for 
special conformal transformations of the argument. 
\par 
Once the homogeneous solutions are known, Section~\ref{sec:S(0,2,z)} works 
out explicitly the MI as given by Euler's formula for the solution of the 
inhomogeneous equation at zeroth order in $(d-2)$, Section~\ref{sec:S(0,4,z)} 
the corresponding zeroth order term of the expansion in $(d-4)$ and 
Section~\ref{sec:S(1,2,z)} the first order term of the expansion in $(d-2)$. 
\par 
Section~\ref{sec:conc} contains the conclusions; 
the first of the two Appendices deals with the relation between the 
``interpolating solutions" and the complete elliptic integrals of the 
first kind, the second Appendix with some definite integrals occurring 
in Sections~~\ref{sec:S(0,2,z)} and \ref{sec:S(1,2,z)}. 
\par 

\section{The differential equations for the MIs.} 
\label{sec:diffeq} 
\begin{figure}[h]
\begin{center}
\includegraphics*[2cm,0cm][10cm,4cm]{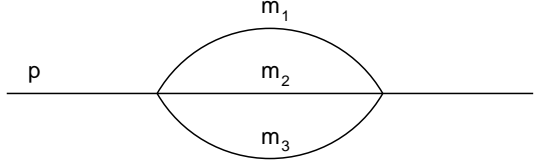}
\caption{\label{fig1} The 2-loop sunrise graph.}
\end{center}
\end{figure}

\setcounter{equation}{0} 
The 2-loop sunrise graph of Fig.~\ref{fig1} with arbitrary masses 
$m_1, m_2, m_3$, is known to possess 4 Master Integrals 
(MI's)~\cite{Tarasov:1997kx}, already extensively studied in the 
literature~\cite{lit}. The system of the 
four linear differential equations in the squared external momentum $p^2$ 
for the Master Integrals (MIs) was written in~\cite{CCLR1}. 
The differential equations were used 
for obtaining analytically specific values or 
behaviours at zero and infinite momentum transfer~\cite{CCLR1}, at 
pseudothresholds~\cite{CCR1} and threshold~\cite{CCR2}, as well 
as for direct numerical integration\cite{CCR3}, but a satisfactory 
knowledge of the analytic expression of the MI's for arbitrary 
momentum transfer and masses is still missing. 
In this paper we tackle that problem in the simpler 
equal masses case. While the algebraic burden 
in the arbitrary mass case will surely be much heavier, there 
are however indications~\cite{HCER} that the approach can be extended to the 
arbitrary mass case as well. \par 
We will work as usual within the $d$-continuous regularization scheme, 
defining the loop integration measure, in agreement with previous work, as 
\begin{equation} 
      {\mathfrak D}^dk = \frac{1}{C(d)} 
                         \frac{d^dk}{(2\pi)^{d-2}} \ , \nonumber 
\end{equation} 
with $C(d)= (4\pi)^{(4-d)/2} \Gamma(3-d/2) $, \ie 
\begin{equation} 
      {\mathfrak D}^dk = \frac{1}{\Gamma\left(3-\frac{d}{2}\right)} 
                         \frac{d^dk}{4\pi^{\frac{d}{2}}} \ . 
\labbel{defDdk} 
\end{equation} 
At $d=4$, it reduces to the standard measure 
\begin{equation} 
       {\mathfrak D}^4k = \frac{d^4k}{(2\pi)^2} \ ; 
\labbel{D4k} 
\end{equation} 
further, with the definition Eq.(\ref{eq:defDdk}), the tadpole reads 
(exactly in $d$) 
\begin{equation} 
  T(d,m) = \int\frac{{\mathfrak D}^dk}{k^2+m^2} 
         = \frac{m^{d-2}}{(d-2)(d-4)} \ . 
\labbel{deftad} 
\end{equation} 
\par 
When all the masses are equal, the 2-loop sunrise has just two MIs; 
putting equal to 1 the value of the mass, the two MI's 
$S(d,p^2)$ and $ S_1(d,p^2)$ can be defined as 
\begin{eqnarray} 
 S(d,p^2) &=& \int \frac{\dnk{1}\ \dnk{2}} 
              {(k_1^2+1)(k_2^2+1)[(p-k_1-k_2)^2+1]} \ , 
  \labbel{defMI} \\ 
 S_1(d,p^2) &=& \int \frac{\dnk{1}\dnk{2}} 
            {(k_1^2+1)^2(k_2^2+1)[(p-k_1-k_2)^2+1]} \ , 
  \labbel{defMI1} 
\end{eqnarray} 
where $p$ is the external momentum. In the following we will use 
\begin{equation} 
    p^2 = z = - u \ , 
\labbel{defz} 
\end{equation} 
with $z$ positive (and $u$ negative) when $p$ is spacelike, and the 
continuation to timelike values ($u$ positive) will be performed with the 
usual replacement 
\begin{equation}
    z = - (u+\iep) \ . 
\labbel{defu}  
\end{equation} 
\par 
The system of linear differential equations in $z$ satisfied by the 
4 MI's in the arbitrary mass case was written in~\cite{CCLR1}; in the 
equal mass limit it goes into the following $2\times 2$ system 
of linear differential equations in $z$ for 
the two MI's defined in Eq.s(\ref{eq:defMI},\ref{eq:defMI1}) 
\begin{eqnarray} 
 z \frac{d}{dz} S(d,z) &=& (d-3) S(d,z) + 3 S_1(d,z) \ , \nonumber\\ 
 z(z+1)(z+9) \frac{d}{dz} S_1(d,z) &=& \frac{1}{2}(d-3)(8-3d)(z+3)S(d,z) 
                                                         \nonumber\\ 
  &+& \frac{1}{2} \left[ (d-4)z^2 + 10(2-d)z +9(8-3d) \right] S_1(d,z) 
                                                         \nonumber\\ 
  &+& \frac{1}{2} \frac{z}{(d-4)^2} \ . 
\labbel{1stosys} 
\end{eqnarray} 
The system can be rewritten as a second order differential equation for 
$S(d,z)$ only 
\begin{eqnarray}
 z(z+1)(z+9) \frac{d^2}{dz^2}S(d,z) && \nonumber\\
  +\ \frac{1}{2}\left[ (12-3d)z^2 + 10(6-d)z + 9d \right] \frac{d}{dz}S(d,z)
                              {\kern3pt} &&  \nonumber\\ 
  +\ \frac{1}{2}(d-3) \left[ (d-4)z - d - 4 \right] S(d,z) 
                              {\kern15pt} && {\kern-18pt} = 
    \frac{3}{2} \frac{1}{(d-4)^2} \ , 
\labbel{2ndoeq} 
\end{eqnarray} 
and the second MI $S_1(d,z)$ can be expressed in terms of $S(d,z)$ and its 
first derivative, 
\begin{equation} 
  S_1(d,z) = \frac{1}{3}\left[ - (d-3) + z\frac{d}{dz} \right] S(d,z) \ . 
\labbel{S1} 
\end{equation} 
From now on we can take $S(d,z)$ and its first derivative 
$dS(d,z)/dz$ as the effective MIs of the problem. 

\section{The expansions in $d$. } 
\label{sec:expansions} 
\setcounter{equation}{0} 
In any physical application, the $d\to4$ limit is to be taken; in that 
limit, $S(d,z)$ can be Laurent-expanded in $(d-4)$ around $d=4$ as 
\begin{equation} 
 S(d,z) = \sum_{n=-2} S^{(n)}(4,z)(d-4)^n \ ,  
\labbel{expat4} 
\end{equation} 
where, as it is well known, the Laurent-series at $d=4$ starts with 
a leading double pole in $(d-4)$. In this paper we will show how to 
evaluate analytically the coefficients $S^{(n)}(4,z)$ up to $n=1$, 
\ie up to the first order term in $(d-4)$ included. 
\par 
Let us recall here that by acting on any Feynman integral in 
$d$-continuous dimensions with a suitable differential operator one 
can obtain an expression relating the value of that integral evaluated 
in $(d-2)$ dimensions to the values, evaluated in $d$ dimensions, of a 
suitable combination of other integrals related to the same Feynman 
graph~\cite{Tarasov}. In the case of 
the Master Integral $S(d,z)$, expressing the result in terms of the 
effective MI's (and of tadpoles) the relation reads 
\begin{eqnarray} 
 S(d-2,z) = \frac{4}{(d-6)^2(z+1)(z+9)} \biggl[ && {\kern-20pt} 
           (d-3)\ \bigl(\; 3(d-2)+2(d-3)z\; \bigr)S(d,z) \biggr. \nonumber\\ 
            &-& 2(d-3)z(z-3)\frac{d}{dz}S(d,z) 
            \;+\; \biggl. \frac{3(z+3)}{4(d-4)^2}\; \biggr] \ . 
\labbel{S(d-2,z)} 
\end{eqnarray}  
By differentiating in $z$ the above relation, and using Eq(\ref{eq:2ndoeq}) 
for eliminating the second derivative of $S(d,z)$, one obtains a similar 
relation for $dS(d-2,z)/dz$. The two relations can be used to express 
$S(d,z)$ in terms of $S(d-2,z)$ and $dS(d-2,z)/dz$; replacing 
further $d$ by $(2+d)$, one finally obtains 
\begin{eqnarray}
 S(2+d,z) &=& \frac{1}{12(d-1)(3d-2)(3d-4)} 
   \biggl\{ 2(d-4)^2(z+1)(z+9) \left[ 1 + (z-3)\frac{d}{dz} \right] S(d,z) 
   \biggr. \nonumber\\ 
   && \biggl. + (d-2)(d-4)^2 (87+22z-z^2) S(d,z) 
     - \frac{36}{(d-2)^2} + \frac{3z-63}{(d-2)} \biggr\} 
\labbel{2+d} 
\end{eqnarray} 
\par 
Quite in general, assume a relation of the form 
\begin{equation} 
   L(2+d) = R(d) 
\end{equation} 
as given; setting $d=2+\eta$ and Laurent-expanding in $\eta$ around 
$d=2$ both the l.h.s., $L(4+\eta) = \sum_n L^{(n)}(4)\eta^n$, and the 
r.h.s., $R(2+\eta) = \sum_n R^{(n)}(2)\eta^n$, one obtains 
\[ \sum_n L^{(n)}(4)\eta^n = \sum_n R^{(n)}(2)\eta^n \ , \] 
which obviously implies, at any order $n$ in the Laurent-expansion 
in $\eta$ 
\[ L^{(n)}(4) = R^{(n)}(2) . \] 

We put $d=2+\eta$ in Eq.(\ref{eq:2+d}) and look for the systematical 
Laurent-expansion of l.h.s. and r.h.s. The l.h.s. is nothing but 
\[ \sum_n S^{(n)}(4,z) \eta^n \ , \] 
where the coefficients $S^{(n)}(4,z)$ are the same as in 
Eq.(\ref{eq:expat4}). Within the r.h.s. of Eq.(\ref{eq:2+d}) we write 
\begin{equation} 
 S(d,z) = \sum_{n=0} S^{(n)}(2,z) \eta^n \ ,  
\labbel{expat2} 
\end{equation} 
where the sum starts from $n=0$ as $S(d,z)$, Eq.(\ref{eq:defMI}), is 
regular at $d=2$; the singularities in $(d-2)$ of the r.h.s. are then 
entirely due to the two last terms in the r.h.s. of Eq.(\ref{eq:2+d}), 
with the double and simple poles in $(d-2)$ coming from the tadpoles 
entering in Eq.(\ref{eq:S(d-2,z)}). That implies in particular that the 
Laurent-expansion of the l.h.s. must also have a double and a simple 
pole, as already anticipated. When expanding also the overall $d$-depending 
coefficient, one finds 
\begin{eqnarray} 
  S^{(-2)}(4,z) &=& - \frac{3}{8} \ , \labbel{S(-2,4)} \\  
  S^{(-1)}(4,z) &=& \frac{1}{32}(z+18) \ , \labbel{S(-1,4)} \\  
   S^{(0)}(4,z) &=& \frac{1}{12}(z+1)(z+9) \left( 
                    1 + (z-3)\frac{d}{dz} \right) S^{(0)}(2,z) \nonumber\\ 
                  && - \frac{1}{128}(72+13z) \ . \labbel{S(0,4)} \\ 
   S^{(1)}(4,z) &=& 
      \frac{1}{12}(z+1)(z+9)\left[ 1 + (z-3)\frac{d}{dz} \right] S^{(1)}(2,z) 
                                                               \nonumber\\ 
 &+& \frac{1}{48}(21-126z-19z^2) S^{(0)}(2,z) \nonumber\\ 
 &-& \frac{17}{48}(z+1)(z+9)(z-3) \frac{dS^{(0)}(2,z)}{dz} \nonumber\\ 
 &+& \frac{5}{512}( 36 + 23z ) \ . 
\labbel{S(1,4)} \ . 
\end{eqnarray} 
Note that the singular part of $S(d,z)$ for $d\to4$ is entirely determined 
by the above equations (and of course in agreement with previous results, 
see for instance Eq.s(53) of~\cite{CCLR1}). \par 
In the following of this paper we will show how to obtain 
$ S^{(0)}(2,z) $ and $ S^{(1)}(2,z) $; $ S^{(0)}(4,z) $ and 
$ S^{(1)}(4,z) $ can then be obtained by the previous formulae. 

\section{The expansion around $d=2$. } 
\label{sec:d=2} 
\setcounter{equation}{0} 
It was shown in the previous Section, 
Eq.(\ref{eq:S(-2,4)}-\ref{eq:S(1,4)}) that the coefficients 
$S^{(n)}(4,z)$ of the expansion of $S(d,z)$ around $d=4$ can be expressed 
in terms of the coefficients $S^{(n)}(2,z)$ of the expansion around $d=2$, 
which is perhaps also simpler, as polar terms in $(d-2)$ are absent. 
In the following we will therefore restrict ourselves to the 
evaluation of the coefficients $S^{(n)}(2,z)$ of the expansion at $d=2$, 
referring to Eq.s(\ref{eq:S(-2,4)}-\ref{eq:S(1,4)}) for obtaining the 
$S^{(n)}(4,z)$. 
\par 
By inserting the expansion Eq.(\ref{eq:expat2}) into Eq.(\ref{eq:2ndoeq}) 
and systematically expanding in $(d-2)$ also all the other $d$-depending 
terms, one obtains a system of chained equations, each equation 
corresponding to a given order in $(d-2)$. All the equations are of the 
form 
\begin{equation} 
  \biggl\{ \ \ \frac{d^2}{dz^2} +
         \left[\frac{1}{z} + \frac{1}{z+1} + \frac{1}{z+9}
        \right] \frac{d}{dz} 
      + \left[ \frac{1}{3z} - \frac{1}{4(z+1)} - \frac{1}{12(z+9)}
         \right] \biggr\} S^{(n)}(2,z) = N^{(n)}(2,z) \ , \labbel{chain} 
\end{equation} 
where $n=0,1,.. $ is the order in the expansion in $(d-2)$. 
Note that the differential operator acting on $S^{(n)}(2,z)$ is the 
same for any value of $n$, while the inhomogeneous terms do depend on 
$n$. The explicit expressions of the first two terms are 
\begin{eqnarray} 
 N^{(0)}(2,z) &=& \frac{1}{24z} - \frac{3}{64(z+1)} + \frac{1}{192(z+9)} 
               = \frac{3}{8z(z+1)(z+9)} \ ,           \nonumber\\ 
 N^{(1)}(2,z) &=& \left( - \frac{1}{2z} + \frac{1}{z+1} + \frac{1}{z+9} 
                  \right)\frac{dS^{(0)}(2,z)}{dz} \labbel{Nchained} \\ 
              &+& \left( \frac{5}{18z} - \frac{1}{8(z+1)} 
                   - \frac{11}{72(z+9)} \right) S^{(0)}(2,z) \nonumber\\ 
              &+& \frac{1}{24z} - \frac{3}{64(z+1)} + \frac{1}{192(z+9)} 
                                                  \ . \nonumber 
\end{eqnarray} 
The system of equations Eq.(\ref{eq:chain},\ref{eq:Nchained}) is chained, 
in the sense that the equation for the coefficient $S^{(n)}(2,z)$ involves 
in general in the inhomogeneous term coefficients $S^{(k)}(2,z)$ (and 
their first derivatives) of lower order $k<n$. Therefore the system must 
to be solved bottom up; as $N^{(0)}(2,z)$ is explicitly known, one 
starts from $n=0$ and obtains $S^{(0)}(2,z)$, 
which then appears as a known term in the inhomogeneous part of the 
equation at $n=1$ for $S^{(1)}(2,z)$, and so on to higher values of $n$. 
\par 
The homogeneous part of Eq.(\ref{eq:chain}), which as already observed 
is the same for any value of $n$, reads 
\begin{equation} 
  \biggl\{ \ \ \frac{d^2}{dz^2} +
         \left[\frac{1}{z} + \frac{1}{z+1} + \frac{1}{z+9}
        \right] \frac{d}{dz} 
      + \left[ \frac{1}{3z} - \frac{1}{4(z+1)} - \frac{1}{12(z+9)}
         \right] \biggr\} \Psi(z) = 0 \ . \labbel{homo} 
\end{equation} 
As it is a second order differential equation, it admits two linearly 
independent solutions, say $\Psi_1,(z), \Psi_2,(z)$; if 
\begin{equation} 
   W(z) = \Psi_1(z) \frac{d\Psi_2(z)}{dz} 
        - \Psi_2(z) \frac{d\Psi_1(z)}{dz} \ , 
\labbel{Wronskian} 
\end{equation} 
is their Wronskian, according to Euler's method of the variation of the 
constants, the solutions of Eq.(\ref{eq:chain}) are given by the integral 
representations 
\begin{eqnarray} 
 S^{(n)}(2,z) &=& \Psi_1(z) \left( \Psi^{(n)}_1 
     - \int_0^z \frac{dw}{W(w)} \Psi_2(w) N^{(n)}(2,w) \right) \nonumber\\ 
       \labbel{Euler} \\ 
              &+& \Psi_2(z) \left( \Psi^{(n)}_2 
     + \int_0^z \frac{dw}{W(w)} \Psi_1(w) N^{(n)}(2,w) \right) \ , \nonumber 
\end{eqnarray} 
where \( \Psi^{(n)}_1, \Psi^{(n)}_2 \) are two integration constants. 
\par 
Up to this point Eq.(\ref{eq:Euler}) is just formal: it will become a 
``substancial", explicit analytic expression only when all the ingredients 
entering in it -- the two $\Psi_i(z)$, their Wronskian and the integration 
constants $ \Psi^{(k)}_i $ -- will be actually evaluated. 
\par 
Although the Wronskian is defined in terms of the $\Psi_i(z)$, it is known 
(Liouville's formula) that it can be immediately obtained (up to a 
multiplicative constant) from 
Eq.(\ref{eq:homo}). An elementary calculation using the definition 
Eq.(\ref{eq:Wronskian}) and the value of the second derivatives of the 
$\Psi_i(z)$, as given by Eq.(\ref{eq:homo}) of which they are solutions, 
leads to the equation 
\[ \frac{d}{dz} W(z) = - \left( \frac{1}{z} + \frac{1}{z+1}
                       + \frac{1}{z+9} \right) W(z) \ , \]
which gives at once 
\begin{equation} 
 W(z) = \frac{9}{z(z+1)(z+9)} \ , \labbel{Wvalue} 
\end{equation} 
where the multiplicative constant has been fixed anticipating later 
results. 
\par 
Evaluating the two $\Psi_i(z)$ (and then the integration constants) 
is much more laborious. 

\section{Building the solution of the homogeneous equation. } 
\label{sec:Building} 
\setcounter{equation}{0} 
Given a differential equation, it is immediate to obtain its solutions 
around any point $z_0$ in the form of a power series in $(z-z_0)$; 
indeed, when the solution is tentatively written as a power series, 
the differential equation can be easily recast as 
a recursive formula for the coefficients of the expansion. 
If the differential equation is of second order, as in Eq.(\ref{eq:homo}), 
the specification of the two initial conditions which identify the solution 
corresponds to the specification of the first two coefficients, needed 
for initializing the recursion. \par 
The expansion converges only up to the nearest singular point; in the 
case of Eq.(\ref{eq:homo}) the singular points are the four points 
$z=0,-1,-9$ and $z=\infty$. 
At each of the four singular points, one finds (see below) that 
Eq.(\ref{eq:homo}) has a regular solution, 
say $\psi_1(z)$, whose series expansion is straightforward, and a 
logarithmically divergent solution; the divergent solution can be written 
as the logarithm of the appropriate argument times the same function 
$\psi_1(z)$ which corresponds to the regular solution, plus another 
regular function, say $\psi_2(z)$, whose series expansion coefficients 
are also recursively provided by the equation. \par 
While obtaining a local solution is immediate, building a solution 
in the whole range $-\infty < z < \infty$ (\ie working out the analytic 
continuation of a local solution) is a much more demanding task. 
At least in principle, one could consider unrelated pairs of 
linearly independent solutions evaluated as series expansions around some 
carefully chosen points, say $\bar{z_j}$, each solution 
depending on a pair of constants (unrelated, to begin with), 
pick up one of the points, say $\bar{z_0}$, 
and start defining the two independent solutions at that point 
by fixing somehow the two constants at that point. 
One can then look at 
the expansion around the nearest point, say $\bar{z_1}$ (it is 
assumed that thanks to the careful choice of the points $\bar{z_j}$ 
the convergence regions of the expansions in $\bar{z_0}$ and 
$\bar{z_1}$ overlap) and fix the two arbitrary constants of the expansion at 
$\bar{z_1}$ (\ie obtain the analytic continuation 
from $\bar{z_0}$ to $\bar{z_1}$) by imposing 
the equality of the solutions in the overlapping region. 
The procedure, which can be implemented without significant loss of 
precision also in a merely numerical way, can then be iterated to cover 
the whole range of interest of $z$. \par 
In order to cover with properly converging expansions also the singular 
points, which are at the boundary of the convergence regions of the 
expansions around nearby points, it is mandatory to 
consider also the expansions around the singular points themselves. 
Therefore, we will first work out the (still unrelated) solutions 
around each of the four singular points as series expansions; 
for joining them into a unique, analytically continued pair of 
solutions, rather than relying on expansions around auxiliary points 
(as in a numerical approach) we will look for ``interpolating solutions", 
valid within the intervals between successive pairs of singular points -- 
namely the three intervals $(0>z>-1)\ ,$ $(-1>z>-9) \ $ 
and $(-9>z>-\infty$) -- 
and then will use the limiting values of the interpolating solution at the 
boundaries of each interval for relating the arbitrary constants 
of the solutions at the singular points, building in such a way 
the desired analytic continuation of the solutions in the whole range of 
the variable $z$. 

\section{The solutions at the singular points. } 
\label{sec:singpoints} 
\setcounter{equation}{0} 
The presence of $1/z$ factors in the coefficients of Eq.(\ref{eq:homo}) 
shows that $z=0$ is a singular point of the equation; when looking for 
a solution whose leading power in $z$ is $z^\alpha$ one finds the 
indicial equation $\alpha^2=0$, indicating that the leading behaviors 
at $z=0$ are $1$ and $\log{z}$, so that around $z=0$ the two independent 
solutions can be written as 
\begin{eqnarray} 
 \Psi_1^{(0)}(z) &=& \psi_1^{(0)}(z) \ , \nonumber\\ 
 \Psi_2^{(0)}(z) &=& \ln{z}\ \psi_1^{(0)}(z) + \psi_2^{(0)}(z)\ . 
\labbel{expat0} 
\end{eqnarray} 
When $\psi_1^{(0)}(z), \psi_2^{(0)}(z) $ are expanded as 
power series in $z$ as 
\begin{equation} 
 \psi_i^{(0)}(z) = \sum_{n=0}^\infty \psi_{i,n}^{(0)}\ z^n \ ; 
                                              \hspace{2cm} i=1,2 
\labbel{psiat0} 
\end{equation} 
Eq.(\ref{eq:homo}) gives for the coefficients $\psi_i^{(0)}(n)$ 
the recursion relations valid for $n \ge 1$ 
\begin{eqnarray} 
 \psi_{1,n}^{(0)} = \frac{1}{9n^2} && \left[ 
                   (-3+10n-10n^2)\psi_{1,n-1}^{(0)}\right. \nonumber\\ 
          &&\left. -(n-1)^2\psi_{1,n-2}^{(0)}\right] \ , \nonumber\\ 
 \psi_{2,n}^{(0)} = \frac{1}{9n^3} && \left[ 
                    (-10n+6)\psi_{1,n-1}^{(0)} \right. \nonumber\\ 
                 && -2(n-1)\psi_{1,n-2}^{(0)}          \nonumber\\ 
                 && -n(10n^2-10n+3) \psi_{2,n-1}^{(0)} \nonumber\\ 
          &&\left.  -n(n-1)^2 \psi_{2,n-2}^{(0)} \right] \ . 
\labbel{psin0rec} 
\end{eqnarray} 
The initial conditions 
\begin{eqnarray} 
  \psi_{i,n}^{(0)} &=& 0 \hspace{5mm} \mbox{\rm if} \hspace{5mm} n < 0 
                                                            \nonumber\\ 
  \psi_{1,0}^{(0)} &=& 1  \nonumber\\ 
  \psi_{2,0}^{(0)} &=& 0 \ , 
\labbel{psin0in} 
\end{eqnarray} 
determine uniquely the coefficients of the expansions. 
For definiteness, the first terms are 
\begin{eqnarray} 
 \psi_1^{(0)}(z) &=& 1 - \frac{1}{3}z + \frac{5}{27}z^2 + ... \nonumber\\ 
 \psi_2^{(0)}(z) &=& - \frac{4}{9}z + \frac{26}{81}z^2 + ... 
\labbel{psin0exp} 
\end{eqnarray} 
The radius of convergence of the series Eq.(\ref{eq:psiat0}) is $1$, as 
the nearby singularity of the solutions of Eq.(\ref{eq:homo}) is 
$z=-1$ or $u=1$, see Eq.(\ref{eq:defu}). 
We were unable to solve the recurrence relations 
Eq.(\ref{eq:psin0rec}) in closed form, but by direct inspection 
one easily see that at large $n$ the coefficients have a $1/n$ 
behaviour (which in turn implies that the radius of convergence is 
$1$). For $ 0<z$, \ie $0>u$ (the Euclidean region), both solutions 
$ \Psi_1^{(0)}(z), \Psi_2^{(0)}(z)$, Eq.(\ref{eq:expat0}) are real; 
as for spacelike $u$ there are no singular points of Eq.(\ref{eq:homo}) 
in the whole range $0>u>-\oo$, both functions can be continued along 
the whole positive $z$-axis $0<z<+\oo$, where they keep real values. 
For $-1<z<0$, if $z=-(u+\iep)$, with $1>u>0$ and $\epsilon>0$, 
$ \Psi_1^{(0)}(z) $ is still real, while 
$$ \ln{z} = \ln(-u-\iep) = \ln{u} - i\pi \ , $$ 
so that $ \Psi_2^{(0)}(-u-\iep) $ develops an imaginary part equal to 
$ - i\pi\psi_1^{(0)}(-u) $. 
\par 
Around $z=-1$, one finds similarly that two independent solutions 
of Eq.(\ref{eq:homo}) are given by 
\begin{eqnarray} 
 \Psi_1^{(1)}(z) &=& \psi_1^{(1)}(z) \ , \nonumber\\ 
 \Psi_2^{(1)}(z) &=& \ln(z+1)\ \psi_1^{(1)}(z) + \psi_2^{(1)}(z)\ , 
\labbel{expat1} 
\end{eqnarray} 
where $\psi_1^{(1)}(z), \psi_2^{(1)}(z) $ are defined by the series 
expansions 
\begin{equation} 
  \psi_i^{(1)}(z) = \sum_{n=0}^\infty \psi_{i,n}^{(1)} (z+1)^n \ , 
                                              \hspace{2cm} i=1,2 
\labbel{psiat1} 
\end{equation} 
with initial conditions 
\begin{eqnarray} 
  \psi_{i,n}^{(1)} &=& 0 \hspace{5mm} \mbox{\rm if} \hspace{5mm} n < 0 
                                                            \nonumber\\ 
  \psi_{1,0}^{(1)} &=& 1  \nonumber\\ 
  \psi_{2,0}^{(1)} &=& 0 \ .  
\labbel{psin1in} 
\end{eqnarray} 
The recursion relations valid for $n \ge 1$ are 
\begin{eqnarray} 
 \psi_{1,n}^{(1)} = \frac{1}{8n^2} && \left[ 
                   (7n^2-7n+2)\psi_{1,n-1}^{(1)} \right. \nonumber\\ 
             &&\left. +(n-1)^2\psi_{1,n-2}^{(1)} \right] \ , \nonumber\\ 
 \psi_{2,n}^{(1)} = \frac{1}{8n^3} && \left[ 
                       (7n-4)\psi_{1,n-1}^{(1)} \right. \nonumber\\ 
                   && +2(n-1)\psi_{1,n-2}^{(1)} \nonumber\\ 
             && +n(7n^2-7n+2)\psi_{2,n-1}^{(1)} \nonumber\\ 
          &&\left. +n(n-1)^2 \psi_{2,n-2}^{(1)} \right] \ . 
\labbel{psin1rec} 
\end{eqnarray} 
For definiteness, the first terms of the expansions are 
\begin{eqnarray} 
 \psi_1^{(1)}(z) &=& 1 + \frac{1}{4}(z+1) + \frac{5}{32}(z+1)^2 + ... 
                                                          \nonumber\\ 
 \psi_2^{(1)}(z) &=& \frac{3}{8}(z+1) + \frac{33}{128}(z+1)^2 + ... 
\labbel{psin1exp} 
\end{eqnarray} 
As in the previous case, it was not possible to solve Eq.s(\ref{eq:psin1rec}) 
in closed form, but the coefficients are easily seen to behave as 
$1/n$ for large $n$, so that the radius of convergence of the series 
Eq.(\ref{eq:psiat1}) is 1, the nearest singularity of the solutions of 
Eq.(\ref{eq:homo}) being at $z=0$. Similarly to the previous 
case, $\Psi_1^{(1)}(z)$ is real in the whole range of convergence of 
the expansion around $z=-1$, while $\Psi_1^{(2)}(z)$ is real for 
$z>-1$ and develops an imaginary part $-i\pi\psi_1^{(1)}(z)$ for 
$z=-(u+\iep)$, $u>1$. \par 
Around $z=-9$ one has similarly 
\begin{eqnarray} 
 \Psi_1^{(9)}(z) &=& \psi_1^{(9)}(z) \ , \nonumber\\ 
 \Psi_2^{(9)}(z) &=& \ln(z+9)\ \psi_1^{(9)}(z) + \psi_2^{(9)}(z)\ , 
\labbel{expat9} 
\end{eqnarray} 
where $\psi_1^{(9)}(z), \psi_2^{(9)}(z) $ are given by the series 
expansions 
\begin{equation} 
   \psi_i^{(9)}(z) = \sum_{n=0}^\infty \psi_{i,n}^{(9)}\ (z+9)^n \ , 
                                              \hspace{1.5cm} i=1,2 
\labbel{psiat9} 
\end{equation} 
with the coefficients $\psi_{i,n}^{(9)}$ determined by the 
initial conditions 
\begin{eqnarray} 
  \psi_{i,n}^{(9)} &=& 0 \hspace{5mm} \mbox{\rm if} \hspace{5mm} n < 0 
                                                            \nonumber\\ 
  \psi_{1,0}^{(9)} &=& 1  \nonumber\\ 
  \psi_{2,0}^{(9)} &=& 0 \ , 
\labbel{psin9in} 
\end{eqnarray} 
and by the recursion relations valid for $n \ge 1$ 
\begin{eqnarray} 
 \psi_{1,n}^{(9)} = \frac{1}{72n^2} && \left[ 
                     (17n^2-17n+6)\psi_{1,n-1}^{(9)} \right. \nonumber\\ 
                && \left. -(n-1)^2\psi_{1,n-2}^{(9)} \right] \ , \nonumber\\ 
 \psi_{2,n}^{(9)} = \frac{1}{72n^3} && \left[ 
                      (17n-12)\psi_{1,n-1}^{(9)} \right.  \nonumber\\ 
                    && -2(n-1)\psi_{1,n-2}^{(9)}          \nonumber\\ 
        &&     +n(17n^2-17n+6)\psi_{2,n-1}^{(9)}          \nonumber\\ 
           &&\left. -n(n-1)^2 \psi_{2,n-2}^{(9)} \right] \ . 
\labbel{psin9rec} 
\end{eqnarray} 
For definiteness, the first terms of the expansions are 
\begin{eqnarray} 
 \psi_1^{(9)}(z) &=& 1 + \frac{1}{12}(z+9) + \frac{7}{864}(z+9)^2 + ... 
                                                          \nonumber\\ 
 \psi_2^{(9)}(z) &=& \frac{5}{72}(z+9) + \frac{97}{10368}(z+9)^2 + ... 
\labbel{psin9exp} 
\end{eqnarray} 
The expected radius of convergence of the series is 8 (the nearest 
singularity is at $z=-1$ or $u=1$), $\Psi_1^{(9)}(z)$ being real 
on the whole interval of convergence, while $\Psi_2^{(9)}(z)$ 
develops an imaginary part $-i\pi\psi_1^{(9)}(z)$ for 
$z=-(u+\iep)$, $u>9$. 
\par 
Finally, for $z\to\infty$ we put $z=1/y$ and find as independent solutions 
\begin{eqnarray} 
 \Psi_1^{(\infty)}(z) &=& y \psi_1^{(\infty)}(y) \ , \nonumber\\ 
 \Psi_2^{(\infty)}(z) &=& y\left( \ln(y)\ \psi_1^{(\infty)}(y) 
                                + \psi_2^{(\infty)}(y) \right) \ , 
\labbel{expatoo} 
\end{eqnarray} 
where $\psi_1^{(\oo)}(y), \psi_2^{(\oo)}(y) $ are the following two 
series in $y$ 
\begin{equation}   
\psi_i^{(\oo)}(y) = \sum_{n=0}^\infty \psi_{i,n}^{(\oo)} y^n \ , 
                                              \hspace{1.5cm} i=1,2 
\labbel{psiatoo} 
\end{equation}   
with the coefficients $\psi_{i,n}^{(\oo)}$ determined by the 
initial conditions 
\begin{eqnarray} 
  \psi_{i,n}^{(\oo)} &=& 0 \hspace{5mm} \mbox{\rm if} \hspace{5mm} n < 0 
                                                            \nonumber\\ 
  \psi_{1,0}^{(\oo)} &=& 1  \nonumber\\ 
  \psi_{2,0}^{(\oo)} &=& 0 \ , 
\labbel{psinooin} 
\end{eqnarray} 
and by the recursion relations valid for $n \ge 1$ 
\begin{eqnarray} 
 \psi_{1,n}^{(\oo)} = - \frac{1}{n^2} && \left[ 
                 (10n^2-10n+3)\psi_{1,n-1}^{(\oo)} \right. \nonumber\\ 
             &&\left.+9(n-1)^2\psi_{1,n-2}^{(\oo)} \right] \ , \nonumber\\ 
 \psi_{2,n}^{(\oo)} = -\frac{1}{n^2} && \left[ 
                      (10-6/n)\psi_{1,n-1}^{(\oo)} \right. \nonumber\\ 
                 && +18(1-1/n)\psi_{1,n-2}^{(\oo)}         \nonumber\\ 
             && +(10n^2-10n+3)\psi_{2,n-1}^{(\oo)}         \nonumber\\ 
            &&\left. +9(n-1)^2\psi_{2,n-2}^{(\oo)} \right] \ . 
\labbel{psinoorec} 
\end{eqnarray} 
For definiteness, the first terms of the expansions are 
\begin{eqnarray} 
 \psi_1^{(\infty)}(z) &=& 1 - 3y + 15y^2 + ...            \nonumber\\ 
 \psi_2^{(\infty)}(z) &=& - 4y + 26y^2 + ... 
\labbel{psinooexp} 
\end{eqnarray} 
\par 
The expected radius of convergence of the series in $y=1/z$ is $1/9$, 
as the nearest singularity is at $u=-z=9$. For $y>0$, \ie $z>0$, 
the two series are real; for $z=-(u+\iep)$, $\infty>u>9$, 
$ \ln{y} = - \ln{u} + i\pi$ and $\Psi_2^{(\infty)}(z)$ develops an 
imaginary part $-i\pi/u\ \psi_1^{(\oo)}(y)$. 

\section{The interpolating solutions. } 
\label{sec:interpol} 
\setcounter{equation}{0} 
Eq.s(\ref{eq:chain}),(\ref{eq:Nchained}) at $n=0$ give the 
equation valid for the $n=2$ limiting value of $S(d,z)$, 
Eq.(\ref{eq:defMI}); as the inhomogeneous term $N^{(0)}(2,z)$, 
Eq.(\ref{eq:Nchained}), is always real for real $z$, the imaginary 
part of $S^{(0)}(2,z)$ satisfies the homogeneous equation 
Eq.(\ref{eq:homo}). As the imaginary part is present only above the physical 
threshold, \ie when $u=-z>9$, it is convenient to rewrite Eq.(\ref{eq:homo}) 
in terms the function $J(u) = \Psi(z)$ and of the variable $u=-z$, 
\begin{equation} 
  \biggl\{ \ \ \frac{d^2}{du^2} +
         \left[\frac{1}{u} + \frac{1}{u-1} + \frac{1}{u-9}
        \right] \frac{d}{du} 
      + \left[ - \frac{1}{3u} + \frac{1}{4(u-1)} + \frac{1}{12(u-9)}
         \right] \biggr\} J(u) = 0 \ . 
\labbel{homou} 
\end{equation} 
As already observed in in~\cite{GP}, 
the Cutkosky-Veltman rule gives for the imaginary part of $S^{(0)}(2,z)$ 
(up to a multiplicative constant irrelevant here), the integral 
representation 
\begin{equation} 
   \int_4^{(\sqrt{u}-1)^2} \frac{db}{\sqrt{R_4(u,b)} } \ , 
\labbel{ImS20} 
\end{equation} 
where $ R_4(u,b) $ stands for the polynomial (of 4th
order in $b$ and 2nd order in $u$)
\begin{equation} 
 R_4(u,b) = b(b-4)(b-(\sqrt{u}-1)^2) (b-(\sqrt{u}+1)^2) \ . 
\labbel{defR4} 
\end{equation} 
For $u>9$ the four roots in $b$ of $R_4(u,b)$ are ordered as 
\begin{equation} 
     0 < 4 < (\sqrt{u}-1)^2 < (\sqrt{u}+1)^2 \ , 
\labbel{rootsofR4} 
\end{equation} 
and the $b$ integration in Eq.(\ref{eq:ImS20}) runs between the two adjacent 
roots $4$ and $(\sqrt{u}-1)^2$. We will now verify that Eq.(\ref{eq:ImS20}) 
does indeed satisfy Eq.(\ref{eq:homou}) by a procedure which will also 
suggest a broader set of similar solutions, also for other ranges of 
values of $u$. 
\par 
To that aim, let us introduce the auxiliary functions 
\begin{equation} 
   H(k,u) = \int_\beta^B \frac{db\ b^k}{\sqrt{R_4(u,b)} } \ , 
\labbel{defHku} 
\end{equation} 
$\beta, B$ are any pair of adjacent roots in $b$ of $R_4(u,b)$, and $k$ 
is an integer (if the integration boundaries do not include the root 
$b=0\ $, $k$ can take any positive or negative value; otherwise, 
$k\ge 0$ for simplicity). One finds immediately 
\[ \int_\beta^B db \frac{d}{db}\left( b^n\sqrt{R_4(u,b)} \right) = 0 \ , \] 
as the integration runs among two points at which the 
primitive of the integrand vanish; 
by working out explicitly the derivatives, moving always the square root 
$\sqrt{R_4(u,b)}$ to the denominator and using the definition 
Eq.(\ref{eq:defHku}) one finds the identity 
\begin{eqnarray} 
 (k+2)H(k+3,u) - (2k+3)(u+3)H(k+2,u) + (k+1)(u+3)^2H(k+1,u) && \nonumber\\ 
               - 2(2k+1)(u-1)^2H(k,u) &=& 0 \ . 
\labbel{Hids} 
\end{eqnarray} 
Another identity is obtained by writing 
$$  \int_\beta^B {\kern-2pt} db \frac{d}{db} \left( 
   \ln{ \frac { b(u-b+3) + \sqrt{R_4(u,b)} } { b(u-b+3) - \sqrt{R_4(u,b)} } } 
                                                         \right) = 0 \ ; $$ 
the derivative of the logarithm is 
$$ \frac{-3b+(u+3)} { \sqrt{R_4(u,b)} } \ , $$ 
from which one finds 
\begin{equation} 
   3H(1,u) - (u+3)H(0,u) = 0 \ . 
\labbel{H10id} 
\end{equation} 
As a consequence of the identities Eq.s(\ref{eq:Hids},\ref{eq:H10id}), 
all the integrals $H(k,u)$, Eq.(\ref{eq:defHku}) can be expressed 
in terms of only two ``master integrals", which can be taken to be 
\begin{eqnarray} 
  H(0,u) &=& \int_\beta^B \frac{db}{\sqrt{R_4(u,b)}} \ , \nonumber\\ 
  H(2,u) &=& \int_\beta^B \frac{db\ b^2}{\sqrt{R_4(u,b)}} \ . 
\labbel{defMHu} 
\end{eqnarray} 
Also the $u$-derivatives of all the $H(k,u)$ can be expressed in terms 
of the two functions. By using Eq.s(\ref{eq:Hids},\ref{eq:H10id}) 
write 
\begin{eqnarray} 
 \int_\beta^B db \sqrt{R_4(u,b)} &=& \left( \frac{3}{2} + \frac{21}{2}u 
       - \frac{3}{2}u^2 + \frac{1}{6}u^3 \right) H(0,u) 
       - \left( \frac{3}{2} + \frac{1}{2}u \right) H(2,u) \nonumber\\ 
 \int_\beta^B db b\ \sqrt{R_4(u,b)} &=& \left( \frac{15}{4} + \frac{61}{4}u 
       + \frac{49}{6}u^2 + \frac{7}{6}u^3 + \frac{1}{12}u^4 
                          + \frac{1}{36}u^5 \right)H(0,u) \nonumber\\ 
     && + \left( - \frac{15}{4} + \frac{3}{4}u - \frac{9}{4}u^2 
                          - \frac{1}{12}u^3 \right)H(2,u) \ . \nonumber 
\end{eqnarray} 
One can then take the $u$-derivatives of both sides of the two equations; 
in the l.h.s. one can differentiate the integrands, and then express the 
result in terms of the two functions $H(0,u), H(2,u)$ times polynomials 
in $u$; in the r.h.s. one finds both the two functions and their 
$u$-derivatives; solving for the derivatives gives 
\begin{eqnarray} 
  \frac{d}{du}H(0,u) &=& \frac{1}{2u(u-1)(u-9)}\left[ 
         ( 3 + 14u - u^2 )H(0,u) - 3 H(2,u) \right]       \nonumber\\ 
  \frac{d}{du}H(2,u) &=& \frac{1}{6u(u-1)(u-9)}\left[ 
                ( 9 + 228u + 30u^2 - 12u^3 + u^4 )H(0,u) \right. \nonumber\\ 
        && {\kern80pt} \left.       + ( -9 - 42u + 3u^2 )H(2,u) \right] \ . 
\labbel{uderH} 
\end{eqnarray} 
With one further differentiation and some algebra one obtains also the 
second derivative 
\begin{eqnarray} 
  \frac{d^2}{du^2}H(0,u) &=& \frac{1}{2u^2(u-1)^2(u-9)^2}\left[ 
     ( - 27 - 12u + 202u^2 - 36u^3 + u^4 ) H(0,u) \right. \nonumber\\ 
   && {\kern92pt} \left.   + (   27 - 60u + 9u^2 )H(2,u) \right] \ . 
\labbel{uder2H} 
\end{eqnarray} 
\par 
One can now substitute $J(u)$ by $H(0,u)$ in Eq.(\ref{eq:homou}); by using 
Eq.s(\ref{eq:uderH},\ref{eq:uder2H}) for expressing the derivatives in 
terms of $H(0,u)$ and $H(2,u)$, one finds that Eq.(\ref{eq:homou}) is 
satisfied. But $H(0,u)$ Eq.(\ref{eq:defMHu}) is equal to the integral 
Eq.(\ref{eq:ImS20}) if $\beta=4$ and $B=(\sqrt{u}-1)^2$; it is so verified 
that Eq.(\ref{eq:ImS20}) satisfy Eq.(\ref{eq:defMHu}), as expected. 
\par 
The verification that $H(0,u)$ satisfies Eq.(\ref{eq:defMHu}) remains 
valid for any choice of the two roots $(\beta,B)$ out of the four 
roots of $R_4(u,b)$ listed in Eq.(\ref{eq:rootsofR4}). The linearly 
independent choices can be taken to be the three real integrals 
\begin{equation} 
 J_1(u) = \int_0^4\frac{db}{\sqrt{-R_4(u,b)}} \ , {\kern18pt} 
 J_2(u) = \int_4^{(\sqrt{u}-1)^2}\frac{db}{\sqrt{R_4(u,b)}} \ , {\kern18pt} 
 J_3(u) = \int_{(\sqrt{u}-1)^2}^{(\sqrt{u}+1)^2} \frac{db}{\sqrt{-R_4(u,b)}} 
                                                 \ , 
\labbel{the3Js} 
\end{equation} 
where the sign in front of $R_4(u,b)$ within the square root 
has been adjusted to make all the 
integrals real (but an overall multiplicative factor $i$ would be 
anyhow irrelevant when dealing with the solutions of a homogeneous 
equation). All the three functions $J_i(u), i=1,2,3$ are solutions 
of Eq.(\ref{eq:homo}), which has however only two linearly independent 
solutions; therefore they cannot be all independent. Indeed, consider 
the contour integral 
\begin{equation} 
   \oint_{(\infty)} db\ \frac{1} {\sqrt{R_4(u,b)}} \ , 
\labbel{defLoo} 
\end{equation} 
where the loop runs on the circle at infinity. For $u>9$ 
the integrand has two cuts; one cut is from $\ b=(\sqrt{u}-1)^2\ $ to 
$\ b=(\sqrt{u}+1)^2$ and correspondingly its value is 
$1/\sqrt{R_4(u,b)} = 1/(i\sqrt{-R_4(u,b)}) $ for $b$ above the cut and 
$ - 1/(i\sqrt{-R_4(u,b)}) $ for $b$ below the cut, the other cut 
from $b=0\ $ to $\ b=4$, and corresponding values 
$ - 1/(i\sqrt{-R_4(u,b)}) $ and $ 1/(i\sqrt{-R_4(u,b)}) $ above and 
below the cut. The integral on the circle 
at infinity (where the integrand behaves as $1/b^2$) vanishes; by 
shrinking the circle to two closed paths around the cuts, one obtains, 
apart an overall factor $2i$, 
\begin{equation} 
  J_1(u) - J_3(u) = 0 \ , 
\labbel{2nd3did}
\end{equation} 
which reduces by one, as expected, the number of the independent solutions. 
\par 
The ordering Eq.(\ref{eq:rootsofR4}) of the four roots in $b$ of 
$R_4(u,b)$ is valid, as already remarked, for $u>9$; when $u$ varies, 
the ordering of the roots varies as well, and the corresponding real 
interpolating solutions of the form of Eq.(\ref{eq:the3Js}) are expressed 
by different choices of the integration interval in the variable $b$. 

\section{The homogeneous solutions. } 
\label{sec:homosol} 
\setcounter{equation}{0} 
We will now use the results of the previous Sections for building two 
independent solutions of Eq.(\ref{eq:homo}), say $\Psi_1(z), \Psi_2(z)$, 
properly continued in the whole range $-\infty<z<\infty$. We start from the 
the point $z=0$; in the interval $1>z>-1$ we define 
\begin{equation} 
  1>z>-1\ \ \left\{ 
  \begin{array}{ll} 
    \mbox{$ \Psi_1(z-\iep) = \Psi_1^{(0)}(z-\iep) \ , $} \\ 
    \\ 
    \mbox{$ \Psi_2(z-\iep) = \Psi_2^{(0)}(z-\iep) \ ,$} 
  \end{array} \right. 
\labbel{Psiat0} 
\end{equation} 
where the two functions $\Psi_1^{(0)}(z),\Psi_2^{(0)}(z)$ are defined by 
Eq.(\ref{eq:expat0}) and the equations following it. 
With that definition, the multiplicative constant of the Wronskian is 
fixed, and the Wronskian takes the value 
\[ W(z) = \frac{9}{z(z+1)(z+9)} \ , \] 
already anticipated in Eq.(\ref{eq:Wvalue}). 
\par 
Eq.(\ref{eq:Psiat0}) is meaningful within the convergence region of 
the series, \ie $ 1>z>-1 $. For positive $z$ the $-\iep$ prescription 
is irrelevant, as the two functions $\Psi_i(z)$ are both 
real; for $0>z>-1,\ $ i.e. $\ 0<u<1,\ \ $ $\Psi_1(z)$ is still real, 
while the factor $\ln(z-\iep)$ present in $\Psi_2(z)$ develops an imaginary 
part $-i\pi$, so that one has 
\begin{equation} 
  0>z>-1\ \ \left\{ 
  \begin{array}{ll} 
    \mbox{$ \Psi_1(z-\iep) = \psi_1^{(0)}(z) \ , $} \\ 
    \\ 
    \mbox{$ \Psi_2(z-\iep) = \ln(-z)\psi_1^{(0)}(z) + \psi_2^{(0)}(z) 
                           - i\pi\psi_1^{(0)}(z) \ ,$} 
  \end{array} \right. 
\labbel{0<Psi<1} 
\end{equation} 
and, quite in general, for real $z$ 
\begin{equation} 
   \Psi_i(z+\iep) = ( \Psi_i(z-\iep) )^* \ . 
\labbel{Psi*} 
\end{equation} 
\par 
In the $z\to0^-$ limit, in particular, one finds 
\begin{eqnarray} 
  \lim_{z\to 0^-} \Psi_1(z-\iep) &=& 1 \ , \nonumber\\ 
  \lim_{z\to 0^-} \Psi_2(z-\iep) &=& \ln(-z) - i\pi \ . 
\labbel{Psiat0-} 
\end{eqnarray} 
\par 
In the interval $0>z>-1$, \ie or $0<u<1$, the four roots of $R_4(u,b)$, 
Eq.(\ref{eq:defR4}), are ordered as 
\[ (0,(\sqrt{u}-1)^2,(\sqrt{u}+1)^2,4) \ ; \] 
according to Section~\ref{sec:interpol}, the interpolating solutions 
can be chosen as 
\begin{eqnarray} 
 J_1^{(0,1)}(u) &=& \int_0^{(\sqrt{u}-1)^2} 
            \frac{db}{\sqrt{-R_4(u,b)}} \ , \nonumber\\ 
 J_2^{(0,1)}(u) &=& \int_{(\sqrt{u}-1)^2}^{(\sqrt{u}+1)^2} 
            \frac{db}{\sqrt{R_4(u,b)}} \ ,  \nonumber\\ 
 J_3^{(0,1)}(u) &=& \int_{(\sqrt{u}+1)^2}^4 
            \frac{db}{\sqrt{-R_4(u,b)}} \ . 
\labbel{Jat01} 
\end{eqnarray} 
The above functions are real, and regular in the region 
$0<u<1$; according to the discussion leading to 
Eq.(\ref{eq:2nd3did}), one has however the identity 
\begin{equation} 
  J_3^{(0,1)}(u) = J_1^{(0,1)}(u) \ , 
\labbel{J013J011} 
\end{equation} 
showing that only two of them can be linearly independent. 
In the $z\to 0^-$ or $u\to 0^+$ limit, the two roots $(\sqrt{u}-1)^2$ and 
$(\sqrt{u}+1)^2$ become equal, so that the integration in $b$ of 
Eq.s(\ref{eq:Jat01}) is elementary and gives 
\begin{eqnarray} 
  \lim_{u\to 0^+} J_1^{(0,1)}(u) &=& \frac{1}{\sqrt{3}} 
                 \left( - \frac{1}{2}\ln{u} + \ln3 \right) \ , \nonumber\\ 
  \lim_{u\to 0^+} J_2^{(0,1)}(u) &=& \frac{\pi}{\sqrt{3}} \ . 
\labbel{Jat0+} 
\end{eqnarray} 
The explicit calculation gives also 
$$ \lim_{u\to 0^+} J_3^{(0,1)}(u) = \frac{1}{\sqrt{3}} 
                \left( - \frac{1}{2}\ln{u} + \ln3 \right) \ 
$$ 
in agreement, of course, with Eq.s(\ref{eq:J013J011}) and (\ref{eq:Jat0+}). 
\par 
Given the different behaviours in the $u\to0^+$ limit, $J_1^{(0,1)}(u)$ 
and $J_2^{(0,1)}(u)$ are linearly independent; in that same limit, one 
finds that their Wronskian 
\[ \lim_{u\to 0^+} \left[ J_1^{(0,1)}(u) \frac{d}{du}J_2^{(0,1)}(u) 
      - \frac{d}{du}J_1^{(0,1)}(u) J_2^{(0,1)}(u) \right] = 
                                             \pi \frac{1}{6} \ , \] 
shows the expected $1/u$ singularity (the $J_i^{(0,1)}(u)$ are 
solutions of Eq.(\ref{eq:homou}), which is equivalent to Eq.(\ref{eq:homo}) 
with $z=-u$ and whose Wronskian Eq.(\ref{eq:Wvalue}) behaves as $1/z$ when 
$z\to0$, up to an overall multiplicative factor). 
\par 
The $\Psi_i(z), i=1,2$, are linearly independent solutions of 
Eq.(\ref{eq:homo}), so that the $J_i^{(0,1)}(u), i=1,2,\ $ which are 
also solutions of the same equation (the equivalence of 
Eq.(\ref{eq:homou}) and Eq.(\ref{eq:homo}) has been repeatedly 
recalled) can be expressed in terms of the $\Psi_i(z)$, and {\it viceversa}. 
By comparing the $z\to 0^-$ behaviour of the $\Psi_i(z)$ given by 
Eq.s(\ref{eq:Psiat0-}) and the $u\to 0^+$ behaviour of the $J_i^{(0,1)}(u)$, 
Eq.s(\ref{eq:Jat0+}), one finds that in the interval $0>z>-1$, 
\ie $0<u<1$ the solutions can be written as 
\begin{eqnarray} 
    u = -z , && 0>z>-1 , {\kern12pt} 0<u<1, \nonumber\\ 
  \Psi_1(z-\iep) &=& \frac{\sqrt{3}}{\pi} J_2^{(0,1)}(u) \ , \nonumber\\ 
  \Psi_2(z-\iep) &=& - 2\sqrt{3}J_1^{(0,1)}(u) 
            + \frac{\sqrt{3}}{\pi}\left( 2\ln{3} - i\pi  \right) 
                                J_2^{(0,1)}(u) \ . 
\labbel{PsitoJ01} 
\end{eqnarray} 
\par 
Similarly, in the $u\to1^-\ $ limit, the two couples of roots 
$(0, (\sqrt{u}-1)^2)$ and $(\ (\sqrt{u}+1)^2,4)$ become equal, so that 
one finds easily 
\begin{eqnarray} 
  \lim_{u\to 1-} J_1^{(0,1)}(u) &=& \lim_{u\to 1-} J_3^{(0,1)}(u) 
                               \ =\  \frac{\pi}{4} \ , \nonumber\\ 
  \lim_{u\to 1-} J_2^{(0,1)}(u) &=& - \frac{3}{4}\ln(1-u) 
                                    + \frac{9}{4}\ln2 \ . 
\labbel{Jat1-} 
\end{eqnarray} 
\par 
But for $u\to1^-\ $ (or $z\to-1^+$) the two interpolating solutions 
$J_i^{(0,1)}(u)$ can also be expressed in terms of the two 
$\Psi_i^{(1)}(z)$, Eq.s(\ref{eq:expat1}); matching the limiting 
values of Eq.s(\ref{eq:Jat1-}) with the corresponding behaviours 
of Eq.s(\ref{eq:expat1}) (the $\Psi_i^{(1)}(z)$ are real for 
$0<z<-1$) gives 
\begin{equation} 
  0>z>-1\ \ \left\{ 
  \begin{array}{ll} 
    \mbox{$ J_1^{(0,1)}(-z) = \frac{\pi}{4} \Psi_1^{(1)}(z) \ , $} \\ 
    \\ 
    \mbox{$ J_2^{(0,1)}(-z) = \frac{9}{4}\ln{2}\ \Psi_1^{(1)}(z)
                     - \frac{3}{4} \Psi_2^{(1)}(z) \ .$} 
  \end{array} \right. 
\labbel{J01toPsi1} 
\end{equation} 

Substituting Eq.s(\ref{eq:J01toPsi1}) 
into Eq.s(\ref{eq:PsitoJ01}), we finally obtain 
\begin{equation} 
  0>z>-2\ \ \left\{ 
  \begin{array}{ll} 
    \mbox{$ \Psi_1(z-\iep) = \frac{9\sqrt{3}}{4\pi} \ln2\Psi_1^{(1)}(z-\iep) 
                   - \frac{3\sqrt{3}}{4\pi} \Psi_2^{(1)}(z-\iep) \ , $} \\ 
    \\ 
    \mbox{$ \Psi_2(z-\iep) = \frac{\sqrt{3}}{4} 
       \left( \frac{18}{\pi}\ln2\ln3 - 2\pi - i9\ln2 \right) 
                                            \Psi_1^{(1)}(z-\iep) $ } \\ 
    \mbox{$ \phantom{\Psi_2(z-\iep)} + \frac{3\sqrt{3}}{4\pi} 
                 \left( - 2\ln3 + i\pi \right) \Psi_2^{(1)}(z-\iep) \ ,$} 
  \end{array} \right. 
\labbel{Psiat1} 
\end{equation} 
which gives the required analytic continuation of the solutions 
$\Psi_i(z)$, as defined by Eq.(\ref{eq:Psiat0}), from the interval 
$1>z>-1$ to the interval $0>z>-2$ (or $0<u<2$), containing the 
singular point $z=-1$ (or $u=1$). 
\par 
The whole procedure can be repeated in the interval $-9>z>-1$ or $1<u<9$. 
The four roots of $R_4(u,b)$, Eq.(\ref{eq:defR4}), are ordered as 
$$(0,\ (\sqrt{u}-1)^2,\ 4,\ (\sqrt{u}+1)^2) \ , $$ 
and the two interpolating solutions are given by 
\begin{eqnarray} 
 J_1^{(1,9)}(u) &=& \int_0^{(\sqrt{u}-1)^2} 
            \frac{db}{\sqrt{-R_4(u,b)}} \ , \nonumber\\ 
 J_2^{(1,9)}(u) &=& \int_{(\sqrt{u}-1)^2}^{4} 
            \frac{db}{\sqrt{R_4(u,b)}} \ , 
\labbel{Jat19} 
\end{eqnarray} 
while for the third solution, according to Eq.(\ref{eq:2nd3did}), one has 
$$ J_3^{(1,9)}(u) = \int_4^{(\sqrt{u}+1)^2} 
            \frac{db}{\sqrt{-R_4(u,b)}} = J_1^{(1,9)}(u) \ . $$ 
\par 
In the $u\to1^+\ $ limit (as in the $u\to1^-\ $ limit) 
the pairs of roots $(0,(\sqrt{u}-1)^2)$ and $(\ (\sqrt{u}+1)^2,4)$ 
become equal and one evaluates easily 
\begin{eqnarray} 
  \lim_{u\to1^+} J_1^{(1,9)}(u) &=& \frac{\pi}{4} \ , \nonumber\\ 
  \lim_{u\to1^+} J_2^{(1,9)}(u) &=& - \frac{3}{4}\ln(u-1) 
                                    + \frac{9}{4}\ln2 \ . 
\labbel{Jat1+} 
\end{eqnarray} 
As the $J_i^{(1,9)}(u)$ are solutions of Eq.(\ref{eq:homou}), 
equivalent to Eq.(\ref{eq:homo}), 
in the interval $1<u<9$ or $-1>z>-9$, they can be written as linear 
combinations of the $ \Psi_i(z-\iep)$; the comparison of 
Eq.s(\ref{eq:Jat1+}) with 
Eq.s(\ref{eq:expat1},\ref{eq:psin1exp}) in the $z\to-1^-\ $ limit 
gives 
\begin{equation} 
  -1>z>-9\ \ \left\{ 
  \begin{array}{ll} 
    \mbox{$ \Psi_1(z-\iep) = \frac{\sqrt{3}}{\pi}\left( J_2^{(1,9)}(u) 
                           + 3iJ_1^{(1,9)}(u) \right) \ , $} \\ 
    \labbel{PsitoJ19} \\       
    \mbox{$ \Psi_2(z-\iep) = \frac{\sqrt{3}}{\pi}\left( 
                 ( \pi + 6i\ln{3} )J_1^{(1,9)}(u) \right. $} \\ 
    \mbox{$ \phantom{\Psi_2(z-\iep) = \frac{\sqrt{3}}{\pi}} \left. 
               + ( 2\ln{3} - i\pi )J_2^{(1,9)}(u) \right) \ . $} 
  \end{array} \right. 
\end{equation} 
\par 
We look now at $u\to9^-$; in that limit, $(\sqrt{u}-1)^2 \to 4,$ and one 
obtains 
\begin{eqnarray} 
  \lim_{u\to9^-} J_1^{(1,9)}(u) &=& \frac{1}{4\sqrt{3}}\left( 
        - \ln(9-u) + 3\ln2 + 2\ln3 \right) \ , \nonumber\\ 
  \lim_{u\to9^-} J_2^{(1,9)}(u) &=& \frac{\pi}{4\sqrt{3}} \ . 
\labbel{Jat9-} 
\end{eqnarray} 
In close analogy to what already done for the interval $0<u<1$, in the 
interval $1<u<9$ the two interpolating solutions $J_i^{(1,9)}(u)$ 
can be expressed in terms of the $\Psi_i^{(9)}(z)$ Eq.s(\ref{eq:expat9}); 
matching the coefficients of the linear combination by using the limiting 
values for $u\to9^-$, one obtains 
\begin{equation} 
  -1>z>-9\ \ \left\{ 
  \begin{array}{ll} 
    \mbox{$ J_1^{(1,9)}(u) = \frac{1}{4\sqrt{3}}\left( 
        (3\ln2 + 2\ln3)\Psi_1^{(9)}(z) - \Psi_2^{(9)}(z) \right) , $} \\ 
    \labbel{J19toPsi9} \\       
    \mbox{$ J_2^{(1,9)}(u) = \frac{\pi}{4\sqrt{3}} \Psi_1^{(9)}(z) \ . $} 
  \end{array} \right. 
\end{equation} 
\par 
When substituting Eq.s(\ref{eq:J19toPsi9}) in Eq.s(\ref{eq:PsitoJ19}) 
one finally obtains 
\begin{equation} 
  -1>z>-17\ \ \left\{ 
  \begin{array}{ll} 
    \mbox{$ \Psi_1(z) = \frac{1}{4\pi} \left[ \pi 
                 + i\ 3(3\ln2+2\ln3) \right] \Psi_1^{(9)}(z) $} \\ 
    \mbox{$ \phantom{\Psi_1(z)} -i\ \frac{3}{4\pi} \Psi_2^{(9)}(z) \ , $}\\ 
    \\ 
    \mbox{$ \Psi_2(z) = \left[ \frac{3}{4}\ln2 + \ln3 
    + \frac{i}{4\pi}(18\ln2\ln3+12\ln^23-\pi^2)\right] \Psi_1^{(9)}(z) $}\\ 
    \mbox{$ \phantom{\Psi_2(z)} - \frac{1}{4\pi}( \pi + i\ 6\ln3 ) 
                                                   \Psi_2^{(9)}(z) \ . $} 
  \end{array} \right. 
\labbel{Psiat9} 
\end{equation} 
The radius of convergence of the expansion around $z=-9$, \ie $u=9$, 
of the $\Psi_i^{(9)}(z)$ is 8, as the nearest singularity is at $u=1$ 
(or $z=-1$), hence the range of validity $-1>z>-17$ or $1<u<17$ of the 
above formula. 
\par 
The last interval to consider is $9<u<\infty$; the four roots of 
$R_4(u,b)$ are ordered as 
$$ 0,4,(\sqrt{u}-1)^2,(\sqrt{u}+1)^2\ \ , $$ 
and two interpolating solutions corresponding to integrating in $b$ 
between two adjacent roots are 
\begin{eqnarray} 
 J_1^{(9,\infty)}(u) &=& \int_0^{4} 
            \frac{db}{\sqrt{-R_4(u,b)}} \ , \nonumber\\ 
 J_2^{(9,\infty)}(u) &=& \int_{4}^{(\sqrt{u}-1)^2} 
            \frac{db}{\sqrt{R_4(u,b)}} \ , 
\labbel{Jat9oo} 
\end{eqnarray} 
while the third solution satisfies 
$$ J_3^{(9,\infty)}(u) = \int_{(\sqrt{u}-1)^2}^{(\sqrt{u}+1)^2} 
            \frac{db}{\sqrt{-R_4(u,b)}} = J_1^{(9,\infty)}(u) \ . $$ 
Proceedings as int the previous case, from the limiting values 
for $u\to9^+$ 
\begin{eqnarray} 
  \lim_{u\to9^+} J_1^{(9,\infty)}(u) &=& \frac{1}{4\sqrt{3}}\bigl( 
        - \ln(u-9) + 3\ln2 + 2\ln3 \bigr) \ , \nonumber\\ 
  \lim_{u\to9^+} J_2^{(9,\infty)}(u) &=& \frac{\pi}{4\sqrt{3}} \ . 
\labbel{Jat9+} 
\end{eqnarray} 
one obtains 
\begin{equation} 
  -9>z>-\infty\ \ \left\{ 
  \begin{array}{ll} 
    \mbox{$ \Psi_1(z-\iep) = \frac{\sqrt{3}}{\pi}\left( -2J_2^{(9,\infty)}(u)
                           + 3iJ_1^{(9,\infty)}(u) \right) \ , $} \\ 
    \labbel{PsitoJ9oo} \\       
    \mbox{$ \Psi_2(z-\iep) = \frac{\sqrt{3}}{\pi}\left( 
                 ( \pi + 6i\ln{3} )J_1^{(9,\infty)}(u) \right. $} \\ 
    \mbox{$ \phantom{\Psi_2(z-\iep) = \frac{\sqrt{3}}{\pi}} \left. 
                         - 4\ln{3} J_2^{(9,\infty)}(u) \right) \ . $} 
  \end{array} \right. 
\end{equation} 
The limiting values for $u\to\infty$ are 
\begin{eqnarray} 
  \lim_{u\to\infty} J_1^{(9,\infty)}(u) &=& \frac{1}{u}\pi \ , \nonumber\\ 
  \lim_{u\to\infty} J_2^{(9,\infty)}(u) &=& \frac{3}{2u}\ln{u} \ , 
\labbel{Jatoo} 
\end{eqnarray} 
implying 
\begin{equation} 
  -9>z>-\infty\ \ \left\{ 
  \begin{array}{ll} 
    \mbox{$ J_1^{(9,\infty)}(u) = - \pi \Psi_1^{(\infty)}(z-\iep) \ , $} \\ 
    \labbel{J9ootoPsioo} \\ 
    \mbox{$ J_2^{(9,\infty)}(u) = - \frac{3}{2}i\pi \Psi_1^{(\infty)}(z-\iep) 
                             + \frac{3}{2} \Psi_2^{(\infty)}(z-\iep) \ . $} 
  \end{array} \right. 
\end{equation} 
Note that according to the definitions of the $\Psi_i^{(\infty)}(z)$, 
Eq.(\ref{eq:expatoo}), the r.h.s. of the above equations are real, as 
expected (according to the definition Eq.(\ref{eq:Jat9oo}), the 
$J_i^{(9,\infty)}(u)$ are indeed real). 
\par 
By substituting Eq.s(\ref{eq:J9ootoPsioo}) into Eq.s(\ref{eq:PsitoJ9oo}) 
one finally obtains, 
\begin{equation} 
  -9>z>-\infty\ \ \left\{ 
  \begin{array}{ll} 
   \mbox{$ \Psi_1(z-\iep) = 
                  - 3\frac{\sqrt{3}}{\pi} \Psi_2^{(\infty)}(z-\iep) \ , $}\\ 
  \labbel{Psiatoo} \\ 
   \mbox{$ \Psi_2(z-\iep) = - {\sqrt{3}}{\pi} \Psi_1^{(\infty)}(z-\iep) 
             - 6\sqrt{3}\frac{\ln3}{\pi} \Psi_2^{(\infty)}(z-\iep) \ . $} \\ 
  \end{array} \right. 
\end{equation} 
The formula holds also in the range $\infty>z>9$ (where the $\iep$ in the 
arguments can be ignored). 

\section{Identities under transformations of the argument. } 
\label{sec:transfids} 
\setcounter{equation}{0} 
Consider the transformation 
\begin{equation} 
       y = \frac{9}{z}\ , \qquad z =  \frac{9}{y} \ . 
\labbel{9/z} 
\end{equation} 
It maps the singular points of Eq.(\ref{eq:homo}), 
$ z=0,-1,-9,\infty $ into themselves, more exactly $ 0 \to \infty \to 0,\ $ 
$ (-1) \to (-9) \to (-1)$. If $z$ is real and negative, $y$ is also real and 
negative, and the $\iep$ prescription is implemented as 
\begin{equation} 
          \frac{9}{z-\iep} = \frac{9}{z} + \iep = y + \iep \ , 
   \qquad \frac{9}{y+\iep} = \frac{9}{y} - \iep = z - \iep \ . 
\labbel{9/zandiep} 
\end{equation} 
As it is easy to verify, if $\Psi(z)$ is a solution of Eq.(\ref{eq:homo}), 
\begin{equation} 
                  \frac{1}{z}\Psi\left(\frac{9}{z}\right) 
\labbel{Psi(9/z)} 
\end{equation} 
is also a solution of the same equation, and can be therefore expressed 
in terms of the solutions $\Psi_1(z)$ and $\Psi_2(z)$ defined in the 
previous Section. The argument applies, of course, to the 
$\Psi_i(z)$ themselves, for which we can therefore write 
\begin{equation} 
  \Psi_i\left(\frac{9}{z}-\iep\right) 
                = \frac{1}{3}\ z \sum_{j=1,2} A_{ij} \Psi_j(z+\iep) \ , 
\labbel{Psi_i(9/z)} 
\end{equation} 
where the coefficients $A_{ij}$ are the elements of a $2\times2$ 
matrix $A$ and the overall factor $1/3$ has been introduced for 
convenience. One can obtain the four coefficients $A_{ij}$ by 
imposing for instance Eq.(\ref{eq:Psi_i(9/z)}) to be valid for 
$z=-1+\eta, y=-9-\eta$, with $\eta\to 0^+$, 
were the limiting values of the $\Psi_i(z)$ are 
known from Eq.s(\ref{eq:expat1},\ref{eq:expat9}). The result is 
\begin{equation} 
  A = \frac{\sqrt3}{\pi}\left( \begin{array}{cc} 
            2\ln3 & -1 \\ 4\ln^2{3}-\frac{1}{3}\pi^2 & -2\ln3 
                               \end{array} \right) \ . \ \ \ 
\labbel{Amatrix} 
\end{equation} 
One can check that with the above values of the $A_{ij}$ 
Eq.(\ref{eq:Psi_i(9/z)}) holds also in the $z\to 0$ limit, when 
the limiting values of the $\Psi_i(z)$ are given by 
Eq.s(\ref{eq:expat0},\ref{eq:expatoo}). 

Eq.(\ref{eq:Psi_i(9/z)}) can also be written, letting $z\to 9/z$, as 
\begin{equation} 
  \Psi_i(z-\iep) = 3 \frac{1}{z} \sum_{j=1,2} A_{ij} 
                  \Psi_j\left(\frac{9}{z}+\iep\right) \ ; 
\nonumber 
\end{equation} 
as the matrix $A$ is real, by taking the complex conjugate and 
recalling Eq.(\ref{eq:Psi*}) one has 
\begin{equation} 
  \Psi_i(z+\iep) = 3\ \frac{1}{z} \sum_{j=1,2} A_{ij} 
                  \Psi_j\left(\frac{9}{z}-\iep\right) \ , \ \ 
\labbel{Psi_i(z/9)} 
\end{equation} 
When chaining Eq.(\ref{eq:Psi_i(z/9)}) and Eq.(\ref{eq:Psi_i(9/z)}) one 
finds the condition 
\begin{equation} 
            A \cdot A = 1 \ , 
\labbel{Atwo=1} 
\end{equation} 
which is of course satisfied by the explicit values given in 
Eq.(\ref{eq:Amatrix}). Eq.(\ref{eq:Atwo=1}) shows also that two of the 
four coefficients $A_{ij}$ are fixed once the other two are given. 
\par 
The transformation Eq.(\ref{eq:9/z}) maps the points $z=-3,3$ into 
themselves. Correspondingly, Eq.s(\ref{eq:Psi_i(9/z)},\ref{eq:Amatrix}) 
give 
\begin{eqnarray} 
 \Psi_2(3) &=& \frac{1}{\sqrt3}( 2\sqrt3\ln3 - \pi )\ \Psi_1(3) \ , 
                                                     \nonumber\\ 
 Re\Psi_2(-3) &=& \frac{1}{\sqrt3}( 2\sqrt3\ln3 + \pi )\ Re\Psi_1(-3) \ , 
                                                     \nonumber\\ 
 Im\Psi_2(-3) &=& \frac{1}{\sqrt3}( 2\sqrt3\ln3 - \pi )\ Im\Psi_1(-3) \ . 
\labbel{Psi2(pm3)} 
\end{eqnarray} 
\vskip0.5cm 
\par 
In terms of the interpolating solutions, Eq.s(\ref{eq:Psi_i(9/z)}) 
become: in the interval $0<u<1$, with $\infty>9/u>9$, 
\begin{eqnarray} 
      0 < &u& < 1                       \nonumber\\ 
   J_1^{(9,\infty)}\left(\frac{9}{u}\right) &=& 
                 u \frac{\sqrt3}{9} J_2^{(0,1)}(u) \ , \nonumber\\ 
   J_2^{(9,\infty)}\left(\frac{9}{u}\right) &=& 
                 u \frac{\sqrt3}{3} J_1^{(0,1)}(u) \ ; 
\labbel{J9oo(9/u)} 
\end{eqnarray} 
in the interval $1<u<9$, with $9>9/u>1$, 
\begin{eqnarray} 
      1 < &u& < 9                       \nonumber\\ 
   J_1^{(1,9)}\left(\frac{9}{u}\right) &=& 
                 u \frac{\sqrt3}{9} J_2^{(1,9)}(u) \ , \nonumber\\ 
   J_2^{(1,9)}\left(\frac{9}{u}\right) &=& 
                 u \frac{\sqrt3}{3} J_1^{(1,9)}(u) \ , 
\labbel{J19(9/u)} 
\end{eqnarray} 
and, in the interval $9<u<\infty$, with $1>9/u>0$, 
\begin{eqnarray} 
      9 < &u& < \infty                   \nonumber\\ 
   J_1^{(0,1)}\left(\frac{9}{u}\right) &=& 
                 u \frac{\sqrt3}{9} J_2^{(9,\infty)}(u) \ , \nonumber\\ 
   J_2^{(0,1)}\left(\frac{9}{u}\right) &=& 
                 u \frac{\sqrt3}{3} J_1^{(9,\infty)}(u) \ . 
\labbel{J01(9/u)} 
\end{eqnarray} 
Eq.s(\ref{eq:J01(9/u)}) are of course equivalent to Eq.s(\ref{eq:J9oo(9/u)}); 
similarly, the second of Eq.s(\ref{eq:J19(9/u)}) is equivalent to the first. 
At $u=3$, in particular, they reduce to 
\begin{equation} 
   J_2^{(1,9)}(3) = \sqrt3 J_1^{(1,9)}(3) \ . 
\labbel{J19(3)} 
\end{equation} 
\par 
We consider next the transformation 
\begin{equation} 
       y = - \frac{z+9}{z+1}\ , \qquad z =  - \frac{y+9}{y+1} \ . 
\labbel{-(z+9)/(z+1)} 
\end{equation} 
which also maps into themselves the singular points of 
Eq.(\ref{eq:homo}), more exactly it maps $ 0 \to (-9) \to 0,\ $ 
and $ (-1) \to \infty \to (-1)$. For real $y, z$ the transformation implies 
for the $\iep$ prescription 
\begin{eqnarray} 
 &&- \frac{z-\iep+9}{z-\iep+1} = - \frac{z+9}{z+1} - \iep = y - \iep \ , 
                               \nonumber\\ 
 &&- \frac{y-\iep+9}{y-\iep+1} = - \frac{y+9}{y+1} - \iep = z - \iep \ . \ 
\labbel{-(z+9)/(z+1)andiep} 
\end{eqnarray} 
One can verify that, if $\Psi(z)$ is a solution of Eq.(\ref{eq:homo}), 
\begin{equation} 
      \frac{1}{z+1} \Psi\left(- \frac{z+9}{z+1}\right) 
\ \ \labbel{Psi(-(z+9)/(z+1))} 
\end{equation} 
is also a solution of the same equation, and therefore a combination 
of $\Psi_1(z)$ and $\Psi_2(z)$. When applied to the $\Psi_i(z)$ themselves 
the argument gives 
\begin{equation} 
  \Psi_i\left(-\frac{z+9}{z+1}-\iep\right) 
        = \frac{\sqrt2}{4}\ (z+1) \sum_{j=1,2} B_{ij} \Psi_j(z-\iep) \ , 
\labbel{Psi_i(-(z+9)/(z+1))} 
\end{equation} 
where the coefficients $B_{ij}$ are the elements of a $2\times2$ 
matrix $B$ and the overall factor $\sqrt2/4$ has been introduced for 
convenience. The $B_{ij}$ can be fixed, for instance, by imposing 
the validity of Eq.(\ref{eq:Psi_i(-(z+9)/(z+1))}) for $z\to0$; the 
result reads 
\begin{equation} 
  B = \frac{\sqrt2}{2\pi}\left( \begin{array}{cc} 
            \pi + i\ 6\ln3 & -i\ 3 \\ 
     4\pi\ln{3} + i( 12 \ln^2{3} - \pi^2 ) & - \pi - i\ 6\ln3 
                               \end{array} \right) \ . \ \ \ 
\labbel{Bmatrix} 
\end{equation} 
One can check that with the above values of the $B_{ij}$ 
Eq.(\ref{eq:Psi_i(-(z+9)/(z+1))}) holds also in the $z\to -1$ limit. 

Substituting $z$ by $ -(z+1)/(z+9) $ Eq.(\ref{eq:Psi_i(-(z+9)/(z+1))}) 
can also be written as 
\begin{equation} 
  \Psi_i(z-\iep) = - \frac{2\sqrt2}{z+1} \sum_{j=1,2} B_{ij} 
                    \Psi_j\left(-\frac{z+9}{z+1}-\iep\right) \ ; 
\labbel{Psi_i(zto-(z+9)/(z+1))} 
\end{equation} 
by chaining Eq.(\ref{eq:Psi_i(-(z+9)/(z+1))}) and 
Eq.(\ref{eq:Psi_i(zto-(z+9)/(z+1))}) one finds 
\begin{equation} 
            B \cdot B = - 1 \ , 
\labbel{Btwo=-1} 
\end{equation} 
satisfied of course by the explicit values Eq.(\ref{eq:Bmatrix}), 
showing again that two of the four coefficients $B_{ij}$ are fixed once 
the other two are given. 

The transformation Eq.(\ref{eq:-(z+9)/(z+1)}) maps $z=3$ into $z=-3$. 
Correspondingly, from Eq.s(\ref{eq:Psi_i(-(z+9)/(z+1))},\ref{eq:Bmatrix}) 
and Eq.(\ref{eq:Psi2(pm3)}) one obtains 
\begin{eqnarray} 
 Re\Psi_1(-3) &=& \Psi_1(3) \ , \nonumber\\ 
 Im\Psi_1(-3) &=& - \sqrt3\Psi_1(3) \ , \nonumber\\ 
 Re\Psi_2(-3) &=& \frac{1}{3}( \sqrt3\pi + 6\ln3 )\Psi_1(3) \ , \nonumber\\ 
 Im\Psi_2(-3) &=& ( \pi - 2\sqrt3\ln3 )\Psi_1(3) \ . 
\labbel{Psi(-3)} 
\end{eqnarray} 
\par 
In terms of the interpolating solutions, in the interval $0<u<1$, with 
$9 < (9-u)/(1-u) < \infty$, Eq.s(\ref{eq:Psi_i(-(z+9)/(z+1))}) 
become 
\begin{eqnarray} 
      0 < &u& < 1                       \nonumber\\ 
   J_1^{(9,\infty)}\left(\frac{9-u}{1-u}\right) &=& 
                  \frac{1-u}{2} J_1^{(0,1)}(u) \ , \nonumber\\ 
   J_2^{(9,\infty)}\left(\frac{9-u}{1-u}\right) &=& 
                  \frac{1-u}{4} J_2^{(0,1)}(u) \ . 
\labbel{J9oo((u-9)/(u-1))} 
\end{eqnarray} 
The range $1<u<9$ is mapped by the transformation into the (spacelike) 
interval $ -\infty < - (9-u)/(u-1) < 0 $, where we did not consider the 
interpolating solutions; the range $9<u<\infty$, finally, is mapped 
into $ 0<(u-9)/(u-1)<1 $, where Eq.s(\ref{eq:Psi_i(-(z+9)/(z+1))}) give 
\begin{eqnarray} 
      9 < &u& < \infty                  \nonumber\\ 
   J_1^{(0,1)}\left(\frac{u-9}{u-1}\right) &=& 
                  \frac{u-1}{4} J_1^{(9,\infty)}(u) \ , \nonumber\\ 
   J_2^{(0,1)}\left(\frac{u-9}{u-1}\right) &=& 
                  \frac{u-1}{2} J_2^{(9,\infty)}(u) \ . 
\labbel{J01((u-9)/(u-1))} 
\end{eqnarray} 
The above relations are of course equivalent to 
Eq.s(\ref{eq:J9oo((u-9)/(u-1))}). 

\vskip0.5cm 

The transformations Eq.s(\ref{eq:9/z},\ref{eq:-(z+9)/(z+1)}) can be 
combined into a third transformation 
\begin{equation} 
   y = - 9\ \frac{z+1}{z+9}\ , \qquad z = -9\ \frac{y+1}{y+9} \ , 
\labbel{-9(z+1)/(z+9)} 
\end{equation} 
which maps the singular points of Eq.(\ref{eq:homo}) into themselves, 
$0\to-1\to0, $ and $-9\to\infty\to-9 $. For real $y,z$ it implies 
for the $\iep$ prescription 
\begin{eqnarray} 
 &&- 9\ \frac{z-\iep+1}{z-\iep+9} = - 9\ \frac{z+1}{z+9} + \iep = y + \iep \ , 
                               \nonumber\\ 
 &&- 9\ \frac{y-\iep+1}{y-\iep+9} = - 9\ \frac{y+1}{y+9} + \iep = z + \iep \ . 
\ \ 
\labbel{-9(z+1)/(z+9)andiep} 
\end{eqnarray} 
Again, if $\Psi(z)$ is a solution of Eq.(\ref{eq:homo}), 
\begin{equation} 
                  \frac{1}{z+9}\Psi\left(-9\frac{z+1}{z+9}\right) 
\labbel{Psi(-9(z+1)/(z+9))} 
\end{equation} 
is also a solution of the same equation, as it is easy to verify, 
and therefore can be expressed in terms of 
$\Psi_1(z)$ and $\Psi_2(z)$ as 
\begin{equation} 
  \Psi_i\left(-9\frac{z+1}{z+9} + \iep \right) = 
        \frac{\sqrt2}{12} (z+9) \sum_{j=1,2} C_{ij}\Psi_j(z-\iep) \ , 
\labbel{Psi_i(-9(z+1)/(z+9))} 
\end{equation} 
where $C$ is a $2\times2$ matrix, and the overall constant 
$\sqrt2/12$ has been introduced for convenience. 
\par 
By writing Eq.(\ref{eq:Psi_i(z/9)}) with $z$ replaced by $-9(z+1)/(z+9)$ 
and then using Eq,(\ref{eq:Psi_i(-(z+9)/(z+1))}) one finds 
\begin{eqnarray} 
   \Psi_i\left(-9\frac{z+1}{z+9} + \iep \right) &=& 
         - \frac{1}{3}\frac{z+9}{z+1} \sum_{j=1,2} A_{ij} 
             \Psi_j\left( -\frac{z+9}{z+1} - \iep \right) \nonumber\\ 
  &=& - \frac{\sqrt2}{12}(z+9) \sum_{j=1,2}\sum_{k=1,2} 
           A_{ij} B_{jk} \Psi_k(z-\iep) \ . 
\labbel{Psi_i(comb)} 
\end{eqnarray} 
By comparison with Eq.(\ref{eq:Psi_i(-9(z+1)/(z+9))}), 
\begin{equation} 
  C = - A \cdot B = 
      \frac{\sqrt2\sqrt3}{2\pi}\left( \begin{array}{cc} 
            2\ln3 - i \pi & -1 \\ 
     4\ln^2{3} + \frac{1}{3}\pi^2 & - 2\ln3- i\ \pi 
                               \end{array} \right) \ . \ \ \ 
\labbel{Cmatrix} 
\end{equation} 
Exchanging $(z-\iep)$ with $-9(z+1)/(z+9)+\iep$, 
Eq.(\ref{eq:Psi_i(-9(z+1)/(z+9))} can also be written as 
\begin{equation} 
  \Psi_i(z-\iep) = \frac{6\sqrt2}{z+9} \sum_{j=1,2} C^*_{ij} 
  \Psi_j\left(-9\frac{z+1}{z+9} + \iep \right) \ . 
\labbel{Psi_i(-9(z+1)/(z+9))b} 
\end{equation} 
Note the appearance of $C^*_{ij}$ in the r.h.s of the previous formula; 
indeed, the sign of $\iep$ in the arguments has changed with respect to 
Eq.(\ref{eq:Psi_i(-9(z+1)/(z+9))}), and according to Eq.(\ref{eq:Psi*}) 
that amounts to take the complex conjugate. By chaining 
Eq.(\ref{eq:Psi_i(-9(z+1)/(z+9))b}) and Eq.(\ref{eq:Psi_i(-9(z+1)/(z+9))}) 
one gets the identity 
\begin{equation} 
          C^* \cdot C = 1 \ , 
\labbel{C*C=1} 
\end{equation} 
which is of course satisfied by the explicit values of the coefficients 
$C_{ij}$ given by Eq.(\ref{eq:Cmatrix}). 
In terms of the interpolating solutions, in the interval $0<u<1$, 
with $1>9(1-u)/(9-u)>0$, Eq.s(\ref{eq:Psi_i(-9(z+1)/(z+9))}) become 
\begin{eqnarray} 
      0 < &u& < 1                       \nonumber\\ 
   J_1^{(0,1)}\left(9\frac{1-u}{9-u}\right) &=& 
                  \frac{\sqrt3}{36}(9-u) J_2^{(0,1)}(u) \ , \nonumber\\ 
   J_2^{(0,1)}\left(9\frac{1-u}{9-u}\right) &=& 
                  \frac{\sqrt3}{6}(9-u) J_1^{(0,1)}(u) \ ; 
\labbel{J01(9(1-u)/(9-u))} 
\end{eqnarray} 
the interval $1<u<9$ is mapped into the spacelike range 
$0>9(1-u)/(9-u)> -\infty$, where the interpolating functions were not 
considered, while for $9<u<\infty$, corresponding to 
$\infty>9(1-u)/(9-u)>9$, one finds the relations 
\begin{eqnarray} 
      9 < &u& < \infty                  \nonumber\\ 
   J_1^{(9,\infty)}\left(9\frac{u-1}{u-9}\right) &=& 
                  \frac{\sqrt3}{18}(u-9) J_2^{(9,\infty)}(u) \ ; \nonumber\\ 
   J_2^{(9,\infty)}\left(9\frac{u-1}{u-9}\right) &=& 
                  \frac{\sqrt3}{12}(u-9) J_1^{(9,\infty)}(u) \ , 
\labbel{J9oo(9(1-u)/(9-u))} 
\end{eqnarray} 
which are of course equivalent to Eq.s(\ref{eq:J01(9(1-u)/(9-u))}). 

\section{The solution $S^{(0)}(2,z)$ at zeroth order in $(d-2)$. } 
\label{sec:S(0,2,z)} 
\setcounter{equation}{0} 
We can now reconsider Eq.(\ref{eq:Euler}) for $n=0$, \ie at zeroth order 
in the expansion in $(d-2)$. With the explicit value of $N^{(0)}(2,z)$ 
given by the first of Eq.s(\ref{eq:Nchained}), Eq.(\ref{eq:Euler}) becomes 
\begin{eqnarray} 
  S^{(0)}(2,z) &=& \Psi_1(z)\left( \Psi_1^{(0)} 
                - \frac{1}{24} \int_0^z dw \ \Psi_2(w) \right) \nonumber\\ 
               &+& \Psi_2(z)\left( \Psi_2^{(0)} 
                + \frac{1}{24} \int_0^z dw \ \Psi_1(w) \right) \ , 
\labbel{S(0,2,z)} 
\end{eqnarray} 
where the functions $\Psi_i(z)$ are known from the previous Sections, 
while the two integration constants are still to be fixed. 
To fix them, we will exploit the information that the solution, 
corresponding to the loop integral Eq.(ref{eq:defMI}), is real 
as far as $u=-z$ is below the physical threshold $u=9$, \ie in the 
whole range $ -\infty<u<9$ or $\infty > z > -9 \ $, with the reality 
condition valid for any $d$, in particular for the zeroth order term of 
the expansion around $d=2$. 
\par 
In the range $0>z>-1$, \ie $0 < u < 1$, using Eq.s(\ref{eq:PsitoJ01}) 
for expressing the $\Psi_i(z)$ in terms of the $J_i^{(0,1)}(u)$, 
the reality condition gives 
\begin{equation} 
 Im \ S^{(0)}(2,z) = - \sqrt{3} J_2^{(0,1)}(u) \ \Psi_2^{(0)} \ , 
\labbel{cond0u1a} 
\end{equation} 
which implies 
\begin{equation} 
   \Psi_2^{(0)} = 0 \ . 
\labbel{cond0u1b} 
\end{equation} 
\par 
In the range $-1>z>-9$, \ie $1<u<9$, using Eq.s(\ref{eq:PsitoJ01}) and 
Eq.s(\ref{eq:PsitoJ19}) one obtains 
\begin{equation} 
 Im \ S^{(0)}(2,z) = \frac{3}{\pi}\ \left( \sqrt{3} \Psi_1^{(0)} 
                   - \frac{1}{4}\int_0^1 dv\ J_1^{(0,1)}(v) \right) 
                   J_1^{(1,9)}(u) \ ; 
\labbel{cond1u9a} 
\end{equation} 
the (qualitative) condition that $S^{(0)}(2,z)$ is real therefore gives 
the (quantitative) result 
\begin{equation} 
  \Psi_1^{(0)} = \frac{\sqrt{3}}{12}\int_0^1dv \ J_1^{(0,1)}(v) \ , 
\labbel{cond1u9b} 
\end{equation} 
or, on account of Eq.(\ref{eq:duJ101b}) of the Appendix, 
\begin{equation} 
  \Psi_1^{(0)} = \frac{\sqrt{3}}{12} \Clpt , 
\labbel{cond1u9c} 
\end{equation} 
where $\Cl_2(\phi)$ is the Clausen function of weight 2. 
\par 
Eq.s(\ref{eq:cond0u1b},\ref{eq:cond1u9c}) determine completely the solution 
Eq.(\ref{eq:S(0,2,z)}), which from now on can be considered as known in 
closed analytic form. Indeed, expressing again, when needed, the $\Psi_i(z)$ 
in terms of the interpolating solutions $J_i(u)$, and on account of the 
results of the Appendix, we can evaluate in particular the behaviour 
of $S^{(0)}(2,z)$ at the singular points of Eq.(\ref{eq:chain}) at $n=0$. 
We find: 
\begin{itemize} 
\item at $z=0$, 
\begin{equation} 
  S^{(0)}(2,0) = \Psi_1^{(0)} 
               = \frac{\sqrt{3}}{12}\int_0^1dv \ J_1^{(0,1)}(v) 
               = \frac{\sqrt{3}}{12} \Clpt \ , 
\labbel{S(2,0,0)} 
\end{equation} 
in agreement with a result of~\cite{Davydychev:1995mq}; 
\item on the mass shell $-z=u=1$ 
\begin{equation} 
  S^{(0)}(2,-1) = \frac{1}{16}\int_0^1dv \ J_2^{(0,1)}(v) 
                = \frac{1}{64}\pi^2 \ ; 
\labbel{S(2,0,-1)} 
\end{equation} 
\item at the threshold $(z+9) \to 0^+$ or $-z=u \to 9^-$ 
\begin{equation} 
  S^{(0)}(2,z) \ \xrightarrow[(z+9) \to 0^+]{\phantom{1}} 
         \ - \frac{\sqrt{3}}{48}\left[ \pi \ln\left(\frac{z+9}{72}\right) 
           + 5\Clpt \right] + {\cal O}\bigl(z+9\bigr) \ ; 
\labbel{S(2,0,-9)} 
\end{equation} 
\item above threshold, with $z=-(u+\iep)$ and  $u>9$, $S^{(0)}(2,z)$ 
develops the imaginary part 
\begin{equation} 
 ImS^{(0)}(2,-u-\iep) = \frac{1}{4}\pi\ J_2^{(9,\infty)}(u) \ , 
\labbel{ImS(2,0,-u)} 
\end{equation} 
in agreement with the comments accompanying the introduction of 
Eq.(\ref{eq:ImS20}) and Eq.(\ref{eq:Jat9oo}); 
\item finally, when $z\to +\infty$ (spacelike region), the behaviour 
of $S^{(0)}(2,z)$ is 
\begin{equation} 
   S^{(0)}(2,z) \ \xrightarrow[z\to +\infty]{\phantom{1}} 
                \ \frac{3}{16z}\ \ln^2{z} 
                + {\cal O}\left(\frac{1}{z^2}\right) \ . 
\labbel{S(2,0,oo)} 
\end{equation} 
\end{itemize} 
\par 
The above results show explicitly that $S^{(0)}(2,z)$ is 
regular at $z=0$ and $z=-1$, as already recalled. 
It can be interesting to observe that by performing the change of variable 
$v \to 9/v$, and then using the first of Eq.s(\ref{eq:J01(9/u)}), 
Eq.(\ref{eq:S(2,0,0)}) becomes 
\begin{equation} 
   S^{(0)}(2,0) = \frac{1}{4} \int_9^\infty \frac{dv}{v} 
                J_2^{(9,\infty)}(v) \ ; 
\labbel{S(2,0,0)oo} 
\end{equation} 
similarly, with the change $v \to (v-9)/(v-1) $ and the second of 
Eq.s(\ref{eq:J01((u-9)/(u-1))}), Eq.(\ref{eq:S(2,0,-1)}) reads 
\begin{equation} 
   S^{(0)}(2,-1) = \frac{1}{4} \int_9^\infty \frac{dv}{v-1} 
                   J_2^{(9,\infty)}(v) \ . 
\labbel{S(2,0,-1)oo} 
\end{equation} 
Quite in general, one can write for $S^{(0)}(2,z)$ the dispersion 
relation 
\begin{equation} 
 S^{(0)}(2,z) = \frac{1}{\pi}\int_9^\infty \frac{dv}{v+z} 
                 ImS^{(0)}(2,-v-\iep) 
              = \frac{1}{4} \int_9^\infty \frac{dv}{v+z} 
                 J_2^{(9,\infty)}(v) \ ,               
\labbel{disprel} 
\end{equation} 
where Eq.(\ref{eq:ImS(2,0,-u)}) has been used; Eq.(\ref{eq:S(2,0,0)oo}) 
and Eq.(\ref{eq:S(2,0,-1)oo}) can then be seen as the dispersion 
relation evaluated at $z=0$ and $z=-1$. 
\par 
We can also easily work out the expansions of $S^{(0)}(2,z)$ around any 
of the singular points points of Eq.(\ref{eq:chain}) at $n=0$, namely 
$z=(0,-1,-9,\infty)$, with the above results providing the initial values 
and using either the known expansions of the 
$\Psi_i(z)$ given in Section~\ref{sec:homosol} or, better, the differential 
equation Eq.(\ref{eq:chain}) for obtaining the coefficients 
of the expansions up to any required order. We find: 
\begin{itemize}  
\item around $z=0$, according to Eq.(\ref{eq:S(2,0,0)}), the 
expansion can be written as 
\begin{equation} 
  S^{(0;0)}(2,z) = \Psi_1^{(0)} + \sum_{k=1,\infty} s_k^{(0)}z^k \ ; 
\labbel{Sansatzat0} 
\end{equation} 
substituting in Eq.(\ref{eq:chain}) with $n=0$, one finds 
\begin{eqnarray} 
  S^{(0;0)}(2,z) &=& \Psi_1^{(0)} \psi_1^{(0)}(z) \nonumber\\ 
  &+& \frac{1}{24}z - \frac{23}{864}z^2 + .... \ , 
\labbel{Sresat0} 
\end{eqnarray} 
where $\psi_1^{(0)}(z)$ is the same as in Eq.(\ref{eq:psiat0}) \ ; 
\item around $z=-1$, according to Eq.(\ref{eq:S(2,0,-1)}), the
expansion can be written as 
\begin{equation}
  S^{(0;1)}(2,z) = \frac{\pi^2}{64} + \sum_{k=1,\infty} s_k^{(1)}(z+1)^k \ , 
\labbel{Sansatzat1} 
\end{equation} 
and the differential equation gives 
\begin{eqnarray} 
  S^{(0;1)}(2,z) &=& \frac{\pi^2}{64} \psi_1^{(1)}(z) \nonumber\\ 
  &-& \frac{3}{64}(z+1) - \frac{3}{128}(z+1)^2 + .... 
\labbel{Sresat1} 
\end{eqnarray} 
where $\psi_1^{(1)}(z)$ is the same as in Eq.(\ref{eq:psiat1}) \ ; 
\item around $z=-9$, according to Eq.(\ref{eq:S(2,0,-9)}), the
expansion can be written as 
\begin{equation}
  S^{(0;9)}(2,z) = \ln(z+9)\ \sum_{k=0,\infty} s_k^{(9)}(z+9)^k 
               + \sum_{k=0,\infty} t_k^{(9)}(z+9)^k \ , 
\labbel{Sansatzat9} 
\end{equation} 
with 
\begin{equation} 
   s_0^{(9)} = - \frac{\sqrt{3}}{48}\pi \ , \hspace{8mm} 
   t_0^{(9)} = - \frac{\sqrt{3}}{48}\left[ -\pi\ln(72) + 5\Clpt \right] \ ; 
\labbel{s09t09} 
\end{equation} 
the differential equation gives 
\begin{eqnarray} 
  S^{(0;9)}(2,z) &=& s_0^{(9)} \left[ \ln(z+9)\ \psi_1^{(9)}(z) + 
                                       \psi_2^{(9)}(z) \right] \nonumber\\ 
               &+& t_0^{(9)}\ \psi_1^{(9)}(z) \nonumber\\ 
               &+& \frac{1}{192}(z+9) + \frac{5}{6912}(z+9)^2 + .... 
\labbel{Sresat9} 
\end{eqnarray} 
where the $\psi_i^{(9)}(z)$ are the same as in Eq.(\ref{eq:psiat9}) \ ; 
\item for large $z$, finally, according to Eq.(\ref{eq:S(2,0,oo)}) the
expansion can be written as
\begin{equation}
  S^{(0;\infty)}(2,z) = \frac{1}{z} \left[ 
           \ln^2{z}\sum_{k=0,\infty} s_k^{(\infty)}\frac{1}{z^k} 
         + \ln{z}  \sum_{k=1,\infty} t_k^{(\infty)}\frac{1}{z^k} 
         +         \sum_{k=1,\infty} u_k^{(\infty)}\frac{1}{z^k} 
                                  \right] \ , 
\labbel{Sansatzatoo} 
\end{equation} 
with 
\[ s_0^{(\infty)} = \frac{3}{16}\ . \] 
Imposing the validity of the differential equation, the expansion 
can be written as 
\begin{eqnarray} 
  S^{(0;\infty)}(2,z) &=& \frac{3}{16z} \left[ 
           \ln^2{z}\ \psi_1^{(\infty)}\left(\frac{1}{z}\right) 
          - 2\ln{z}\ \psi_2^{(\infty)}\left(\frac{1}{z}\right) 
                                                       \right. \nonumber\\ 
        &+& \left. \frac{2}{z} - \frac{1}{2z^2} + .... \right] \ , 
\labbel{Sresatoo} 
\end{eqnarray} 
where the $\psi_i^{(\infty)}(z)$ are the same as in 
Eq.(\ref{eq:psiatoo}). 

\end{itemize}  

The correspoding expansions for $dS^{(0)}(2,z)/dz$ is immediately 
obtained by differentiation. From Eq.(\ref{eq:S1}) one can then obtain 
the $z$-expansions for $S_1^{(0)}(2,z)$, the zeroth order term in $(d-2)$ 
of the master integral $S_1(d,z)$ Eq.(\ref{eq:defMI1}). 
\par 
The above $z$-expansions can be used as the starting building blocks for 
implementing a computer routine for the fast and precise numerical 
evaluation of $S^{(0)}(2,z)$. 

\section{The solution $S^{(0)}(4,z)$ at zeroth order in $(d-4)$. } 
\label{sec:S(0,4,z)} 
\setcounter{equation}{0} 
From the results of the previous Section 
and from Eq.(\ref{eq:S(0,4)}) one immediately obtains 
the relevant $z$-expansions for $S^{(0)}(4,z)$, 
the zeroth order term in $(d-4)$ of $S(d,z)$ Eq.(\ref{eq:defMI}): 
\begin{itemize} 
\item around $z=0$ 
\begin{eqnarray} 
  S^{(0;0)}(4,z) &=& \Psi_1^{(0)} \left( \frac{3}{2} + \frac{1}{3}z 
                   - \frac{1}{27}z^2 + .... \right) \nonumber\\ 
  &-& \frac{21}{32} - \frac{3}{128}z + \frac{11}{1728}z^2 + .... \ , 
\labbel{S4resat0} 
\end{eqnarray} 
with $\Psi_1^{(0)}$ given by Eq.(\ref{eq:cond1u9c}); 
\item around $z=-1$ 
\begin{eqnarray} 
  S^{(0;1)}(4,z) &=& \frac{\pi^2}{64} \left( - \frac{1}{2}(z+1)^2 
                  - \frac{5}{8}(z+1)^3 + .... \right) \nonumber\\ 
  &-& \frac{59}{128} + \frac{3}{128}(z+1) + \frac{5}{64}(z+1)^2 
   + \frac{37}{384}(z+1)^3 + .... \ ; 
\labbel{S4resat1} 
\end{eqnarray} 
\item around $z=-9$ 
\begin{eqnarray} 
  S^{(0;9)}(4,z) &=& \biggl( s_0^{(9)}\ln(z+9) + t_0^{(9)} \biggr) 
                  \left( \frac{1}{54}(z+9)^2 + \frac{1}{648}(z+9)^3 
                                                 + ... \right)  \nonumber\\ 
  &+& s_0^{(9)} \left( 8 - \frac{4}{9}(z+9) - \frac{4}{9}(z+9)^2 
                         + \frac{7}{5832}(z+9)^3 + ... \right)  \nonumber\\ 
  &+& \frac{45}{128} - \frac{23}{384}(z+9) - \frac{1}{1728}(z+9)^2 
                        + \frac{1}{10368}(z+9)^3 + ... \ , 
\labbel{S4resat9} 
\end{eqnarray} 
with $s_0^{(9)}$ and $t_0^{(9)}$ as in Eq.(\ref{eq:s09t09}); 
\item for large and positive $z$ (spacelike $u$), finally 
\begin{eqnarray} 
  S^{(0;\infty)}(4,z) &=& \ln^2{z}\left( \frac{3}{32} + \frac{3}{16z} 
                               - \frac{3}{16z^2} + ... \right)  \nonumber\\ 
  &+& \ln{z}\left( \frac{1}{32}z + \frac{9}{32z} + \frac{3}{8z^2} 
                                                 + ... \right)  \nonumber\\ 
  &-& \frac{13}{128}z - \frac{15}{32} - \frac{3}{64z} 
                                                 + \frac{29}{64z^2} + ... 
\labbel{S4resatoo} 
\end{eqnarray} 

\end{itemize} 
The above expansions can be used as the starting building blocks for the 
fast and precise numerical evaluation of $S^{(0)}(4,z)$. 

\section{The solution $S^{(1)}(2,z)$ at first order in $(d-2)$. } 
\label{sec:S(1,2,z)} 
\setcounter{equation}{0} 
Once $S^{(0)}(2,z)$, is known, $N^{(1)}(2,z)$, given by the second of 
Eq.s(\ref{eq:Nchained}), is also known, so that we can 
work out Eq.(\ref{eq:chain}) and its solution Eq.(\ref{eq:Euler}) 
for $n=1$, \ie at first order in the expansion in $(d-2)$. 
After some algebra, it takes the relatively simple expression 
\begin{eqnarray} 
 S^{(1)}(2,z) &=& \Psi_1(z)\biggl[ \Psi_1^{(1)} 
      + \Psi_1^{(0)} \left( - \ln3 - \frac{1}{4}\ln{z} 
         + \frac{1}{2} \ln(z+1) + \frac{1}{2} \ln(z+9) \right) 
         \nonumber\\ && {\kern26pt} 
    -  \frac{1}{96}\int_0^z dw\left( 4 - \frac{1}{w} + \frac{2}{w+1} 
               + \frac{2}{w+9} \right) \Psi_2(w) \biggr] \nonumber\\ 
   &+& \Psi_2(z)\biggl[ \Psi_2^{(1)} + \frac{1}{4} \Psi_1^{(0)} \nonumber\\ 
   && {\kern26pt} 
    + \frac{1}{96}\int_0^z dw\left( 4 - \frac{1}{w} + \frac{2}{w+1} 
               + \frac{2}{w+9} \right) \Psi_1(w) \biggr] \ , 
\labbel{S(1,2,z)} 
\end{eqnarray} 
where $\Psi_1^{(0)}$ is given by Eq.(\ref{eq:cond1u9c}) (according to 
Eq.(\ref{eq:cond0u1b}) $\Psi_2^{(0)}=0$) and the two integration constants 
$\Psi_i^{(1)}$ are still to be determined. Imposing as in the previous 
Section the reality conditions for $\infty>z>-9$, we find 
\begin{equation} 
   \Psi_2^{(1)} = 0 \ , 
\labbel{Psi2(1)} 
\end{equation} 
and 
\begin{eqnarray} 
  \Psi_1^{(1)} &=& \frac{\sqrt{3}}{24}\int_0^1 dv 
               \biggl[ \frac{\pi}{3} J_2^{(0,1)}(v) 
      + \left( 2 + \ln{3} + \frac{1}{2}\ln{v} - \ln(1-v) - \ln(9-v) 
        \right) J_1^{(0,1)}(v) \biggr] \nonumber\\ 
    &=& \frac{\sqrt{3}}{12} \biggl[ \Clpt - \frac{3}{4}\Lstpvt 
         + \frac{1}{2}\ln{3}\Clpt - \frac{1}{24}\pi^3 \biggr] \ , 
\labbel{Psi2(2)} 
\end{eqnarray} 
where use is made of the definite integrals of the Appendix. 
\par 
$S^{(1)}(2,z)$ is now fully determined in closed analytic form. Expressing 
again the $\Psi_i(z)$ in terms of the interpolating solutions 
$J_i(u)$ and on account of the results of the Appendix, we obtain 
in particular the following behaviours: 
\begin{itemize} 
\item at $z=0$ 
\begin{equation} 
 S^{(1)}(2,0) = \Psi_1^{(1)} = 
\frac{\sqrt{3}}{12} \biggl[ \Clpt - \frac{3}{4}\Lstpvt   
         + \frac{1}{2}\ln{3}\Clpt - \frac{1}{24}\pi^3 \biggr] \ , 
\labbel{S(2,1,0)} 
\end{equation} 
in agreement with~\cite{Davydychev:1995mq} if allowance is made for 
the difference in our integration measure Eq.(\ref{eq:defDdk}), which 
generates the extra term $\Cl(\pi/3)$; 
\item on the mass shell $-z=u=1$ 
\begin{equation} 
  S^{(1)}(2,-1) = \frac{1}{64}\left( \pi^2 + 3\pi^2\ln{2} 
                            - \frac{21}{2}\zeta(3) \right) \ ; 
\labbel{S(2,1,-1)} 
\end{equation} 
\item at the threshold $(z+9) \to 0^+$ or $-z=u \to 9^-$ 
\begin{eqnarray} 
  S^{(1)}(2,z) \ \xrightarrow[(z+9) \to 0^+]{\phantom{1}} 
  &-& \sqrt{3}\frac{\pi}{96}\ln\left(\frac{z+9}{72}\right) 
             \biggl[ \ln(z+9) + 3\ln{2} - \ln{3} + 2 \biggr] \nonumber\\ 
  &+& \sqrt{3}\frac{5}{64} \biggl[ \Lstpvt 
            - \frac{2}{3}(2+6\ln{2}+\ln{3})\Clpt \biggr]     \nonumber\\ 
  &+& \sqrt{3}\frac{1}{48}\beta_3 + \sqrt{3}\frac{17}{3456}\pi^3 
                             + {\cal O}\bigl(z+9\bigr) \ ; 
\labbel{S(2,1,-9)} 
\end{eqnarray} 
\item finally, when $z\to +\infty$ (spacelike region), the behaviour 
of $S^{(1)}(2,z)$ is 
\begin{equation} 
   S^{(1)}(2,z) \ \xrightarrow[z\to +\infty]{\phantom{1}} 
                \ \frac{1}{32z}\biggl[ 3\ln^3{z} + 6\ln^2{z} 
                - \pi^2\ln{z} + 18\zeta(3) \biggr] 
                + {\cal O}\left(\frac{1}{z^2}\right) \ . 
\labbel{S(2,1,oo)} 
\end{equation} 
\end{itemize} 

\par 
The corresponding complete expansions in $z$ for $S^{(1)}(2,z)$, 
$dS^{(1)}(2,z)/dz$, $S^{(1)}(4,z)$ etc. can be obtained as in the case of 
$S^{(0)}(2,z)$; we omit them for the sake of brevity. Let us just observe 
that from Eq.(\ref{eq:S(1,4)}) and Eq.(\ref{eq:S(2,1,-1)}) one gets 
\begin{equation} 
  S^{(1)}(4,-1)\ =\ \frac{8}{3}S^{(0)}(2,-1) + \frac{65}{512} 
               \ =\ \frac{\pi^2}{24} + \frac{65}{512} \ , 
labbel{S(1,4,-1)} 
\end{equation} 
in agreement with the known result~\cite{forinst}. 

\section{ Conclusions. } 
\label{sec:conc} 
\setcounter{equation}{0} 
We have considered the system of first order differential equations 
in the external 
momentum transfer satisfied by the Master Integrals of the two loop sunrise 
graph in the equal mass limit and in the usual continuous $d$-dimensional 
regularisation. The system is equivalent to a single second order equation 
for the scalar MI; after recalling the relation between the expansions 
in $(d-4)$ and $(d-2)$, the second order equation is expanded around 
$d=2$ and solved by means of the variation of the constants formula by 
Euler. That requires the knowledge of the solutions of the homogeneous 
equation, which is obtained by working out first the 
series expansions of the differential equation at all its singular points, 
then joining smoothly those expansions by means of a set of ``interpolating 
solutions", whose explicit form (one-dimensional definite integrals of a 
suitable integrand) is suggested by the Cutkosky-Veltman rule for the 
imaginary part of the MI. The interpolating solutions, whose 
relation to the complete elliptic integrals of the first kind is 
discussed in an Appendix, are 
found to transform on themselves for a particular set of   
transformations of the arguments. 
Once the full knowledge of the homogeneous equation is established, 
Euler's formula provides in closed analytic form also the limiting values 
of the solutions at all the singular points of the equation. They are 
worked out explicitly for the zeroth order in $(d-2)$, the 
related zeroth order in $(d-4)$ and the first order in $(d-2)$. 
\par 
The analytic behaviour at the singular points and the related expansions can 
be used as the starting building blocks for the implementation of a fast 
and precise numerical routine for the evaluation of the equal-mass sunrise 
scalar integral for arbitrary values of the momentum transfer. The extension 
of the approach to other two-loop self-mass amplitudes with a single mass 
scale seems easy to obtain; in that way one could complete the analytic 
and numerical evaluation of the two-loop electron self-mass in QED. 
While such a result is somewhat academical, much more interesting, 
even if correspondingly difficult (but not out of reach) will be the 
extension of the approach to the two-loop self-mass integrals in the 
general mass case, as required by the precision tests of the current 
Standard Model (QCD \& EW interactions) of elementary particles. 

\section*{Acknowledgement} 
\noindent
We are grateful to J. Vermaseren for his kind assistance in the use
of the algebra manipulating program {\tt FORM}~\cite{FORM}, by which
all our calculations were carried out.

\begin{appendix} 
\renewcommand{\theequation}{\mbox{\Alph{section}.\arabic{equation}}} 

\section{ Relations with the Complete Elliptic Integrals. } 
\label{sec:elliptic} 
\setcounter{equation}{0} 
The integral Eq.(\ref{eq:ImS20}), which is also the phase 
space of three particles of equal mass, can be written as a Complete 
Elliptic Integral of the First Kind~\cite{GP}, (the first explicit 
reference to phase space integrals and elliptic integrals was 
perhaps made in~\cite{Almgren}); the same is true of course for all the other 
integrals corresponding to the interpolating solutions. \par 
Quite in general, given a polynomial of fourth order in $b$, written as 
\begin{equation} 
      R_4(b) = (b-b_1)(b-b_2)(b-b_3)(b-b_4) \ , 
\labbel{R4b1} 
\end{equation} 
where the four roots are ordered as 
\begin{equation} 
      b_1 < b_2 < b_3 < b_4 \ , 
\labbel{orderbi} 
\end{equation} 
and the integral 
\begin{equation} 
   I(b_1,b_2) = \int_{b_1}^{b_2} \frac{db}{\sqrt{-R_4(b)}} \ , 
\labbel{defIb1b2} 
\end{equation} 
performing the standard change of variable~\cite{GR} 
\begin{equation} 
   x^2 = \frac{(b_4-b_2)(b-b_1)}{(b_2-b_1)(b_4-b)} \ , \hspace{6mm}
   b = \frac{b_4(b_2-b_1)x^2+b_1(b_4-b_2)}{(b_1-b_1)x^2+(b_4-b_2)} \ , 
\labbel{btox2} 
\end{equation} 
one obtains 
\begin{equation} 
   I(b_1,b_2) = \int_{b_1}^{b_2} \frac{db}{\sqrt{-R_4(b)}} 
              = \frac{2}{\sqrt{(b_4-b_2)(b_3-b_1)}} K(m) \ , 
\labbel{I12toKm} 
\end{equation} 
where 
\begin{equation} 
   m = \frac{(b_2-b_1)(b_4-b_3)}{(b_4-b_2)(b_3-b_1)} 
\labbel{defm} 
\end{equation} 
and $K(m)$ is the Complete Elliptic Integral of the First Kind, 
\begin{equation} 
   K(m) = \int_0^1\frac{dx}{\sqrt{(1-x^2)(1-mx^2)}} \ . 
\labbel{defK(m)}
\end{equation}
Similarly, one finds 
\begin{equation} 
   I(b_2,b_3) = \int_{b_2}^{b_3} \frac{db}{\sqrt{R_4(b)}} 
              = \frac{2}{\sqrt{(b_4-b_2)(b_3-b_1)}} K(1-m) \ , 
\labbel{I23toK} 
\end{equation} 
where $m$ is the same as in Eq.(\ref{eq:defm}), 
and finally 
\begin{equation} 
   I(b_3,b_4) = \int_{b_3}^{b_4} \frac{db}{\sqrt{-R_4(b)}} 
              = \frac{2}{\sqrt{(b_4-b_2)(b_3-b_1)}} K(m) \ , 
\labbel{I34toK} 
\end{equation} 
so that, quite in general 
\begin{equation} 
   I(b_1,b_2) = I(b_3,b_4) \ , 
\labbel{I12=I34} 
\end{equation} 
in agreement with Eq.s(\ref{eq:the3Js},\ref{eq:2nd3did}). 
\par 
By using the above results, we can express the interpolating solutions 
introduced in Section~\ref{sec:homosol} in terms of complete elliptic 
integrals. We find: 
\begin{itemize} 
\item for the interpolating solutions in the interval $0<u<1$, 
defined by Eq.s(\ref{eq:Jat01}), 
\begin{eqnarray} 
  J_1^{(0,1)}(u) &=& \frac{1}{\sqrt{(1+\ru)^3(3-\ru)}} K\bigl(a(u)\bigr) 
                                                  \ , \nonumber\\ 
  J_2^{(0,1)}(u) &=& \frac{1}{\sqrt{(1+\ru)^3(3-\ru)}} K\bigl(1-a(u)\bigr) 
                                                  \ , \nonumber\\ 
  a(u) &=& \frac{(1-\ru)^3(3+\ru)}{(1+\ru)^3(3-\ru)} \ ; 
\labbel{J01toK} 
\end{eqnarray} 
\item for the interpolating solutions in the interval $1<u<9$, 
defined by Eq.s(\ref{eq:Jat19}), 
\begin{eqnarray} 
  J_1^{(1,9)}(u) &=& \frac{1}{\sqrt{16\ru}} K\bigl(b(u)\bigr) 
                                                  \ , \nonumber\\ 
  J_2^{(1,9)}(u) &=& \frac{1}{\sqrt{16\ru}} K\bigl(1-b(u)\bigr) 
                                                  \ , \nonumber\\ 
  b(u) &=& \frac{(\ru-1)^3(3+\ru)}{16\ru} \ ; 
\labbel{J19toK} 
\end{eqnarray} 
\item finally, for the interpolating solutions in the interval $9<u<\infty$, 
defined by Eq.s(\ref{eq:Jat9oo}), 
\begin{eqnarray} 
  J_1^{(9,\infty)}(u) &=& \frac{1}{\sqrt{(\ru-1)^(\ru+3)}} K\bigl(c(u)\bigr) 
                                                  \ , \nonumber\\ 
  J_2^{(9,\infty)}(u) &=& \frac{1}{\sqrt{(\ru-1)^(\ru+3)}} 
                              K\bigl(1-c(u)\bigr) \ , \nonumber\\ 
  c(u) &=& \frac{16\ru}{(\ru-1)^3(3+\ru)} \ . 
\labbel{J9ootoK} 
\end{eqnarray} 
\end{itemize} 
\par 
The identities under transformations of the argument, written in 
Section~\ref{sec:transfids} for the interpolating solutions, can be 
rewritten in terms of the elliptic integrals. We obtain, in terms of 
the arguments $a(u),b(u),c(u)$ defined in 
Eq.s(\ref{eq:J01toK},\ref{eq:J19toK},\ref{eq:J9ootoK}): 
\begin{itemize} 
\item from Eq.s(\ref{eq:J9oo(9/u)}) or Eq.s(\ref{eq:J01(9/u)}) 
\begin{equation} 
  \frac{K\bigl(1-a(u)\bigr)}{K\bigl(a(u)\bigr)} = 3 
  \frac{K\bigl(  c(t)\bigr)} 
       {K\bigl(1-c(t)\bigr)} \ , 
\labbel{J01toK(9/u)} 
\end{equation} 
with $t=9/u$; 
\item from Eq.s(\ref{eq:J19(9/u)}) 
\begin{equation} 
  \frac{K\bigl(1-b(u)\bigr)}{K\bigl(b(u)\bigr)} = 3 
  \frac{K\bigl(  b(t)\bigr)} 
       {K\bigl(1-b(t)\bigr)} \ , 
\labbel{J19toK(9/u)} 
\end{equation} 
again with $t=9/u$; 
\item from Eq.s(\ref{eq:J01((u-9)/(u-1))}) or 
Eq.s(\ref{eq:J9oo((u-9)/(u-1))}) 
\begin{equation} 
  \frac{K\bigl(1-a(u)\bigr)}{K\bigl(a(u)\bigr)} = 2 
  \frac{K\bigl(1-c(v)\bigr)} 
       {K\bigl(  c(v)\bigr)} \ , 
\labbel{J01toK((u-9)/(u-1))} 
\end{equation} 
with $v = (u-9)/(u-1)$; 
\item from Eq.s(\ref{eq:J01(9(1-u)/(9-u))}) 
\begin{equation} 
  \frac{K\bigl(1-a(u)\bigr)}{K\bigl(a(u)\bigr)} = 6 
  \frac{K\bigl(  a(w)\bigr)} 
       {K\bigl(1-a(w)\bigr)} \ , 
\labbel{J01toK(9(1-u)/(9-u))} 
\end{equation} 
with $w = 9(u-1)/(u-9)$; 
\item finally from Eq.s(\ref{eq:J9oo(9(1-u)/(9-u))}) 
\begin{equation} 
  \frac{K\bigl(1-c(u)\bigr)}{K\bigl(c(u)\bigr)} = \frac{3}{2} 
  \frac{K\bigl(  c(w)\bigr)} 
       {K\bigl(1-c(w)\bigr)} \ . 
\labbel{J9ootoK(9(1-u)/(9-u))} 
\end{equation} 
again with $w = 9(u-1)/(u-9)$. 
\end{itemize} 
\par 
Eq.(\ref{eq:J01toK(9/u)}) is based on the algebraic transformation 
$a(u) \to 1-c(9/u)$; when the parameter $u$ is in the range 
$0<u<1$ all the arguments are real and in the range $(0,1)$, and 
all the elliptic functions take real and positive values 
(the equation remains of course valid, by analytic continuation, 
for any value of $u$). 
Similarly, all the arguments are in the range $(0,1)$ 
in Eq.s(\ref{eq:J01toK((u-9)/(u-1))}) and 
(\ref{eq:J01toK(9(1-u)/(9-u))}) also for $0<u<1$, 
in Eq.(\ref{eq:J19toK(9/u)}) for $1<u<9$ 
and for $9<u<\infty$ in Eq.(\ref{eq:J9ootoK(9(1-u)/(9-u))}), but the 
relations can be continued to arbitrary values of $u$. 
The algebraic relation derived from Eq.(\ref{eq:J01toK(9/u)}) 
by eliminating the parameter $u$ between $a(u)$ and $1-c(9/u)$ is a 
modular equation of degree 3~\cite{Weisstein}; similarly, 
Eq.(\ref{eq:J19toK(9/u)}) 
gives a modular equation also of degree 3 between $b(u)$ and 
$1-b(9/u)$, Eq.(\ref{eq:J01toK((u-9)/(u-1))}) 
a modular equation of degree 2 between $a(u)$ and $c((u-9)/(u-1))$, 
Eq.(\ref{eq:J01toK(9(1-u)/(9-u))}) an equation of degree 6 between $a(u)$ 
and $1-a(9(1-u)/(9-u))$ and finally Eq.(\ref{eq:J9ootoK(9(1-u)/(9-u))}) 
a modular equation of degree $3/2$ between $c(u)$ and $1-c(9(1-u)/(9-u))$.

\section{Definite Integrals. } 
\label{sec:AA} 
\setcounter{equation}{0} 
In this appendix we discuss the evaluation of the definite integrals 
used in the previous Sections. As a first case, we consider 
\begin{equation} 
 \int_0^1 du \ J_1^{(0,1)}(u) = \int_0^1 du \int_0^{(1-\sqrt{u})^2} 
                            \frac{db}{\sqrt{-R_4(u,b)}} \ , 
\labbel{duJ101} 
\end{equation} 
where $J_1^{(0,1)}(u)$ is defined by Eq.(\ref{eq:Jat01}) and $R_4(u,b)$ 
by Eq.(\ref{eq:defR4}). One can invert the order of integration, 
obtaining 
\begin{eqnarray} 
 \int_0^1 du \ J_1^{(0,1)}(u) &=& \int_0^1 \frac{db}{\sqrt{b(4-b)}} 
      \int_0^{(1-\sqrt{b})^2} 
      \frac{du}{\sqrt{ ((1-\sqrt{b})^2-u) ((1+\sqrt{b})^2-u) }} \nonumber\\ 
      &=& - \frac{1}{2}\int_0^1 \frac{db}{\sqrt{b(4-b)}} \ln{b} \ , 
\labbel{duJ101a} 
\end{eqnarray} 
where the $u$-integration is trivial. One can then perform the standard 
change of variable 
\begin{equation} 
  b = \frac{(1+t)^2}{t} \ , 
\labbel{btot} 
\end{equation} 
which for $0<b<4$ is inverted as 
\begin{equation} 
  t = \frac{ \sqrt{b}-i\sqrt{4-b} }{ \sqrt{b}+i\sqrt{4-b} } \ , 
\labbel{ttob} 
\end{equation} 
with $t$ varying in the unit circle between $t=-1$ at $b=0$ and 
$t=\e^{-2i\pi/3}$ at $b=1$; one finds 
\begin{eqnarray} 
 \int_0^1 du \ J_1^{(0,1)}(u) &=& \frac{i}{2}\int_{-1}^{\e^{-2i\pi/3}} 
    \frac{dt}{t}\bigl(\ 2\ln(1+t) - \ln{t}\ \bigr) \nonumber\\ 
 &=& i\biggl[ -\Li_2(-t) - \frac{1}{4}\ln^2{t} \biggr]_{-1}^{\e^{-2i\pi/3}} 
                                                \nonumber\\ 
 &=& \Cl_2\left(\frac{\pi}{3}\right) \ , 
\labbel{duJ101b} 
\end{eqnarray} 
where use has been made of the formula 
$$ \Li_2(\e^{i\phi}) = \frac{1}{6}\pi^2 - \frac{1}{2}\pi\phi 
                     + \frac{1}{4}\phi^2 \ + i\Cl_2(\phi) , $$ 
and 
\begin{equation} 
    \Cl_2(\phi) = - \int_0^\phi d\theta \ln\left(2\sin\frac{\theta}{2}\right) 
\labbel{defCl} 
\end{equation} 
is the Clausen function of weight 2. 
\par 
Along the same lines, one easily obtains 
\begin{eqnarray} 
  \int_0^1 du \ J_2^{(0,1)}(u) &=& \int_0^4 \frac{db}{ \sqrt{b(4-b)} } 
           \int_{(1-\sqrt{b})^2}^1 \frac{du}{\sqrt{ ((1+\sqrt{b})^2-u) 
                    (u-(1-\sqrt{b})^2) } } \nonumber\\ 
    &=& \frac{i}{2} \int_0^4 \frac{db}{ \sqrt{b(4-b)} } 
     \ln\frac{ \sqrt{b}-i\sqrt{4-b} }{ \sqrt{b}+i\sqrt{4-b} } 
    \nonumber\\  &=& \frac{1}{4}\pi^2 \ , 
\labbel{duJ201} 
\end{eqnarray} 
and, as expected from Eq.(\ref{eq:J013J011}), 
$$ \int_0^1 du \ J_3^{(0,1)}(u) = \Cl_2\left(\frac{\pi}{3}\right) \ . $$ 
\par 
The same approach gives also at once 
\begin{equation} 
\int_1^9 du \ J_2^{(1,9)}(u) = \frac{3}{4}\pi^2 \ . 
 \labbel{duJ219} 
\end{equation} 
The case of $J_1^{(1,9)}(u)$ is more complicated; indeed, 
following the same procedure as for the previous integrals one finds 
\begin{eqnarray} 
  \int_1^9 du \ J_1^{(1,9)}(u) &=& \int_0^4 \frac{db}{ \sqrt{b(4-b)} } 
           \int_{(\sqrt{b}+1)^2}^9 \frac{du}{\sqrt{ (u-(1+\sqrt{b})^2) 
                    (u-(1-\sqrt{b})^2) } } \nonumber\\ 
    &=& \frac{1}{2} \int_0^4 \frac{db}{ \sqrt{b(4-b)} } 
     \ln\frac{ 8-b+\sqrt{(4-b)(16-b)} }{ 8-b-\sqrt{(4-b)(16-b)} } \ , 
\labbel{duJ119} 
\end{eqnarray} 
where the new square root $\sqrt{16-b}$ has appeared. 
If one uses at this point the change of variable Eq.s(\ref{eq:btot}), 
as a consequence of the factor $\sqrt{16-b}$ in the argument of the 
logarithm a new quadratic square root, $\sqrt{1-14t+t^2}$, appears; 
that square root can be in turn removed by a suitable change of 
variable, and the result can be expressed, by brute force, as a combination 
of many dilogarithms of unusual and rather nasty arguments. It can be more 
convenient to consider the auxiliary integral 
\begin{equation} 
  A(m) = \int_0^4 \frac{db}{ \sqrt{b(4-b)} } 
           \int_{(\sqrt{b}+m)^2}^{(2+m)^2} \frac{du} 
                {\sqrt{ (u-(m+\sqrt{b})^2) (u-(m-\sqrt{b})^2) } } \ . 
\labbel{auxduJ119m} 
\end{equation} 
Taking the derivative with respect of $m^2$ of the integral representation 
for $A(m)$ gives 
\[ \frac{d}{dm^2}A(m) = - \frac{i}{2m^2} 
                          \int_0^4 \frac{db}{ \sqrt{b(b-4(m+1)^2)} } 
                        = - \frac{1}{2m^2} ( - 2\ln{x} - i\pi ) \ , \] 
where $x$ is defined through $m+1 = 2x/(x^2+1)$. The primitive in $x$ 
of that expression is 
\begin{eqnarray} 
       A(m(x)) &=& i\biggl[ 2i\pi\ln(1-x) - i\pi\ln(1+x^2) 
                        + 4\ln{x}\ln(1-x) + 4\Li_2(x) \nonumber\\ 
               && - \ln(x^2)\ln(1+x^2) - \Li_2(-x^2) 
                                       + \frac{1}{4}\pi^2 \biggr] \ , 
\labbel{auxduJ119ma} 
\end{eqnarray} 
where the constant term $\pi^2/4$ is fixed by observing that $A(m=-1)$ 
is equal to the integral Eq.(\ref{eq:duJ201}) times $i$. At $m=1$, 
or $x=\e^{-i\pi/3}$, Eq.(\ref{eq:auxduJ119ma}) gives 
\begin{equation} 
  \int_1^9 du \ J_1^{(1,9)}(u) = 5 \Cl_2\left(\frac{\pi}{3}\right) \ .  
\labbel{duJ119r}
\end{equation} 
The same result can be obtained by using the identity 
Eq.(\ref{eq:J19(9/u)}); by the change of variable $u=9/v$ one has 
\begin{equation} 
 \int_1^9 du \ J_1^{(1,9)}(u) = \frac{\sqrt{3}}{9}\int_1^9 dv 
                                     \ \frac{1}{v} J_2^{(1,9)}(v) \ , 
\labbel{duJ119s} 
\end{equation} 
and the integration can be continued, as for Eq.(\ref{eq:duJ219}), 
along the line followed for Eq.s(\ref{eq:duJ101a},\ref{eq:duJ201}). 
\par 
With a suitable combinations of the above described techniques, one 
obtains also, with $U>>1$ 
\begin{eqnarray} 
 \int_9^U du \ J_1^{(9,\infty)}(u) &=& \pi\ln{U} - 5\Clpt \nonumber\\ 
 \int_9^U du \ J_2^{(9,\infty)}(u) &=& \frac{3}{4}\ln^2{U} 
                                                 - \frac{1}{4}\pi^2 \ . 
\label{duJ9oo} 
\end{eqnarray} 
\par 
The extension of the above formulae to the case in which a logarithmic 
factor is also present in the integrand requires much additional effort. 
We list the results 
\begin{eqnarray} 
  \int_0^1 du \ J_1^{(0,1)}(u) \ln{u} &=& 3\Lstpvt + \frac{1}{6}\pi^3 \ , 
                                                          \nonumber\\ 
  \int_0^1 du \ J_2^{(0,1)}(u) \ln{u} &=& 2\pi\Clpt - 7\zeta(3) \ , 
                                                          \nonumber\\ 
  \int_0^1 du \ J_1^{(0,1)}(u) \ln(1-u) &=& -\frac{1}{36}\pi^3 \ , 
                                                          \nonumber\\ 
  \int_0^1 du \ J_2^{(0,1)}(u) \ln(1-u) &=& -\frac{21}{8}\zeta(3) \ , 
                                                          \nonumber\\ 
  \int_0^1 du \ J_1^{(0,1)}(u) \ln(9-u) &=& 3\Lstpvt + \frac{5}{18}\pi^3 \ , 
                                                          \nonumber\\ 
  \int_0^1 du \ J_2^{(0,1)}(u) \ln(9-u) &=& \frac{35}{8}\zeta(3) \ , 
\labbel{duJ01log} 
\end{eqnarray} 
\begin{eqnarray} 
  \int_1^9 du \ J_1^{(1,9)}(u) \ln{u} &=& 15\Lstpvt + \frac{23}{18}\pi^3 \ , 
                                                          \nonumber\\ 
  \int_1^9 du \ J_2^{(1,9)}(u) \ln{u} &=& 7\zeta(3) \ , 
                                                          \nonumber\\ 
  \int_1^9 du \ J_1^{(1,9)}(u) \ln(u-1) &=& \beta_3 \ , 
                                                          \nonumber\\ 
  \int_1^9 du \ J_2^{(1,9)}(u) \ln(u-1) &=& \frac{21}{8}\zeta(3) \ , 
                                                          \nonumber\\ 
  \int_1^9 du \ J_1^{(1,9)}(u) \ln(9-u) &=& 15\Lstpvt + \frac{37}{36}\pi^3 
                                            + \beta_3 \ , \nonumber\\ 
  \int_1^9 du \ J_2^{(1,9)}(u) \ln(9-u) &=& 5\pi\Clpt - \frac{35}{8}\zeta(3) 
                                                      \ , 
\labbel{duJ19log} 
\end{eqnarray} 
\begin{eqnarray} 
  \int_9^U du \ J_1^{(9,\infty)}(u) \ln{u} &=& \frac{1}{2}\pi\ln^2{U} 
                 - \frac{10}{9}\pi^3 - 15\Lstpvt      \ , \nonumber\\ 
  \int_9^U du \ J_2^{(9,\infty)}(u) \ln{u} &=& \frac{1}{2}\ln^3{U} - \zeta(3) 
                                                      \ , \nonumber\\ 
  \int_9^U du \ J_1^{(9,\infty)}(u) \ln(u-1) &=& \frac{1}{2}\pi\ln^2{U} 
                        + \frac{1}{12}\pi^3 - \beta_3 \ , \nonumber\\ 
  \int_9^U du \ J_2^{(9,\infty)}(u) \ln(u-1) &=& \frac{1}{2}\ln^3{U} 
                                - \frac{3}{2}\zeta(3) \ , \nonumber\\ 
  \int_9^U du \ J_1^{(9,\infty)}(u) \ln(u-9) &=& \frac{1}{2}\pi\ln^2{U} 
           - \frac{43}{36}\pi^3 - 15\Lstpvt - \beta_3 \ , \nonumber\\ 
  \int_9^U du \ J_2^{(9,\infty)}(u) \ln(u-9) &=& \frac{1}{2}\ln^3{U} 
              + \frac{11}{2}\zeta(3) - 5\pi\Clpt      \ . 
\labbel{duJ9oolog} 
\end{eqnarray} 
The constants $\Lstpvt$ and $\beta_3$ appearing in the above equations 
are defined as 
\begin{eqnarray} 
 \Ls_3(\phi) &=& 
  - \int_0^\phi d\theta \ln^2\left(2\sin\frac{\theta}{2}\right) 
                                                    \ , \nonumber\\ 
 \beta_3 &=& - \frac{3}{4}\pi\ln^2{3} 
             - \frac{3}{2}\pi\Li_2\left(-\frac{1}{3}\right) 
             - 18\ Im \hbox{S}_{12}(i\sqrt{3}) \ . 
\labbel{defLandb} 
\end{eqnarray} 
\par 
As a final remark, in obtaining the previous results we obtained also the 
relations 
\begin{eqnarray} 
   \hbox{Cl}_2\left(\frac{\pi}{6}\right) &=& \frac{2}{3}G 
                                  + \frac{1}{4}\Clpt \ , \nonumber\\  
   \hbox{Ls}_3(\phi) + \hbox{Ls}_3(\pi-\phi) &=& - \frac{1}{6}\pi^3 \ , 
\labbel{newids?} 
\end{eqnarray} 
where $G$ is the Catalan constant, which we could not find in the 
literature. 

\end{appendix} 

\newpage 
\def\NP{{\sl Nucl. Phys.}} 
\def\PL{{\sl Phys. Lett.}} 
\def\PR{{\sl Phys. Rev.}} 
\def\PRL{{\sl Phys. Rev. Lett.}} 
\def\NC{{\sl Nuovo Cim.}}
\def\APP{{\sl Acta Phys. Pol.}}
\def\ZP{{\sl Z. Phys.}}
\def\MPL{{\sl Mod. Phys. Lett.}} 
\def\EPJ{{\sl Eur. Phys. J.}} 
\def\IJMP{{\sl Int. J. Mod. Phys.}}

\end{document}